\documentclass[acmsmall,screen]{acmart}
\usepackage[utf8]{inputenc}

\usepackage{multirow,makecell}
\usepackage{tcolorbox}
\usepackage{color,xcolor}
\usepackage{listings,amsfonts}
\usepackage{mathtools}
\usepackage{colortbl}
\usepackage{tikz}
\usepackage{caption}
\usepackage{subcaption}
\usepackage{threeparttable}
\usepackage{bbding}
\usepackage{graphicx}
\usepackage{textcomp}
\usepackage{paralist}
\usepackage{graphicx}

\usepackage{newunicodechar}
\newunicodechar{₁}{$_1$}
\newunicodechar{≤}{\leq{}}
\usepackage{fancyhdr}
\usepackage{booktabs}
\usepackage{url}
\usepackage{newunicodechar}
\usepackage{pdflscape}
\usepackage{listings}
\usepackage{tabularx}
\usepackage{pifont}

\usepackage{float}
\usepackage{enumitem}

\tcbset{
  colback=gray!20,
  colframe=gray!80,
  boxrule=0.5mm,
  arc=1mm,
  boxsep=1.2mm,
  left=2mm,right=2mm,top=2mm,bottom=2mm,
}

\newtcolorbox{principlebox}[1][]{
  width=\linewidth,
  title={#1},
  fonttitle=\bfseries\footnotesize\ttfamily,
  fontupper=\small,
}

\definecolor{dkgreen}{rgb}{0,0.6,0}
\definecolor{gray}{rgb}{0.4,0.4,0.4}
\definecolor{mauve}{rgb}{0.58,0,0.82}
\definecolor{darkblue}{rgb}{0.0,0.0,0.6}
\definecolor{lightblue}{rgb}{0.0,0.0,0.9}
\definecolor{cyan}{rgb}{0.0,0.6,0.6}
\definecolor{darkred}{rgb}{0.6,0.0,0.0}

\definecolor{yellow}{RGB}{255,255,153}
\definecolor{grey}{RGB}{220,220,220}
\definecolor{green}{RGB}{0,100,0}

\definecolor{KWCo
lor}{rgb}{0.37,0.08,0.25}
\definecolor{CommentColor}{rgb}{0.133,0.545,0.133}
\definecolor{StringColor}{rgb}{0,0.126,0.941}
\definecolor{commentgreen}{RGB}{2,112,10}
\definecolor{eminence}{RGB}{108,48,130}
\definecolor{weborange}{RGB}{255,165,0}
\definecolor{frenchplum}{RGB}{129,20,83}

\AtBeginDocument{%
  \providecommand\BibTeX{{%
    \normalfont B\kern-0.5em{\scshape i\kern-0.25em b}\kern-0.8em\TeX}}}

\setcopyright{acmcopyright}
\copyrightyear{2025}
\acmYear{2025}
\acmDOI{XXXXXXX.XXXXXXX}

\acmJournal{TOSEM}
\acmVolume{X}
\acmNumber{Y}
\acmArticle{1}
\acmMonth{9}

\begin{document}
\title{Can Large Language Models Detect Real-World Android Software Compliance Violations?}

\author[H Zhang]{Haoyi Zhang}
\email{hyeliozhang@gmail.com}
\affiliation{%
  \institution{Xi'an Jiaotong-Liverpool University}
  \city{SuZhou}           
  \country{China}
}

\author[H Ran]{Huaijin Ran}
\email{seventeen17510@gmail.com}
\affiliation{%
  \institution{Xi'an Jiaotong-Liverpool University}
  \city{SuZhou}           
  \country{China}
}

\author[X Tang]{Xunzhu Tang}
\authornote{Xunzhu Tang is the corresponding author (xunzhu.tang@uni.lu).}
\email{xunzhu.tang@uni.lu}
\affiliation{%
  \institution{University of Luxembourg}
  \city{Luxembourg}     
  \country{Luxembourg}
}

\begin{abstract}
The rapid development of Large Language Models (LLMs) has transformed software engineering, showing promise in tasks like code generation, bug detection, and compliance checking. However, current models struggle to detect compliance violations in Android applications across diverse legal frameworks. We propose \emph{CompliBench}, a novel evaluation framework for assessing LLMs' ability to detect compliance violations under regulations like LGPD, PDPA, and PIPEDA. The framework defines two tasks: Task 1 evaluates \emph{retrieval and localization} at file, module, and line granularities, and Task 2 assesses \emph{multi-label judgment} for code snippets. These tasks mirror the audit process, where auditors locate problematic code and determine implicated provisions. Traditional metrics fail to capture important aspects like cross-granularity stability and jurisdictional consistency. Thus, we introduce stability-aware composites (SGS, RCS, CRGS, and OCS) for a more comprehensive assessment. Experiments with six models, including GPT-4O and Claude-3.5, show \emph{CompliBench} improves compliance detection, with Claude-3.5-sonnet-20241022 achieving the highest OCS score (0.3295), and Gemini-2.5-pro the lowest (0.0538). This work demonstrates \emph{CompliBench}'s potential for improving LLM performance in compliance tasks and provides a foundation for future tools aligned with data protection standards. Our project is available at \url{https://github.com/Haoyi-Zhang/CompliBench}.
\end{abstract}

\maketitle




\section{Introduction}
\label{sec:introduction}

Large language models (LLMs) are increasingly employed in software engineering (SE) tasks, particularly those involving code search, summarization, and repository triage. However, most existing SE benchmarks emphasize code \emph{generation} or single-file reasoning (e.g., HumanEval, MBPP, CodeXGLUE, SWE-bench) \cite{chen2021humaneval,austin2021mbpp,lu2021codexglue,jimenez2023swe-bench}. These frameworks do not address practical challenges faced by compliance reviewers when auditing mobile applications. Specifically, questions such as: \emph{Where in this Android project does the code collect identifiers or transmit telemetry? And which statutory articles does that behavior implicate?} These questions require not only repository-scale retrieval (ranging from file to module to line) but also statute-aware judgment, going well beyond simply generating patches for small snippets.

Over the past decade, the Android privacy community has developed robust static, dynamic, and hybrid analyses—such as FlowDroid, Amandroid, IccTA, and TaintDroid—to track sensitive data flows and detect potential violations \cite{arzt2014flowdroid,wei2014amandroid,li2015iccta,enck2010taintdroid}. Meanwhile, legal and policy NLP approaches have advanced the textual analysis of privacy commitments, such as OPP-115 and Polisis \cite{wilson2016opp115,harkous2018polisis}. However, comprehensive evaluations linking concrete repository evidence to specific statutory articles remain scarce. While GDPR has received considerable attention in policy-text analysis, other data protection laws, including Brazil's LGPD, Singapore's PDPA, and Canada's PIPEDA, have been underrepresented in benchmarks that directly link legal reasoning to code.

\textbf{CompliBench} addresses this gap by constructing a code-grounded benchmark for three under-studied statutes: LGPD, PDPA, and PIPEDA. Unlike previous benchmarks centered on GDPR, CompliBench offers article-specific, repository-linked evaluations, directly connecting code evidence to statutory articles. It preserves auditor-grade provenance, including commit hashes, stable file paths, and explicit anchors at the file/module/line level, ensuring transparency and traceability in the evaluation process.

\paragraph{Tasks that mirror auditor workflows.} \textit{CompliBench} defines two complementary tasks that reflect real-world compliance reviewer workflows for Android projects (Java/Kotlin/XML). \textbf{Task~1} focuses on cross-granularity \emph{retrieval}: Given a suspected scenario, can a system surface the relevant \emph{files}, \emph{modules}, or even \emph{line}-level anchors? To evaluate retrieval performance, we report six standard IR metrics: Precision@k, R-Precision, Mean Reciprocal Rank (MRR), Mean Average Precision (MAP), normalized Discounted Cumulative Gain (nDCG), and Accuracy@k/HitRate. These metrics are commonly used in information retrieval research \cite{manning2008ir, jarvelin2002ndcg, herlocker2004evaluating}. \textbf{Task~2} targets fragment-level \emph{judgment}: Given a curated code snippet, can a system predict the full set of applicable statutory articles? We evaluate judgment performance using six multi-label classification metrics: Micro-F1, Macro-F1, Jaccard index, Hamming loss, Subset Accuracy (exact match), and normalized Coverage Error, all of which are well-established in multi-label classification tasks \cite{zhang2014multilabel, hamming1950, tsoumakas2010mlsurvey, schapire2000boostexter}.

The raw dataset offers balanced coverage across jurisdictions and granularities:
\emph{LGPD} (File/Module/Line/Snippet = 90/90/208/208), \emph{PDPA} (77/78/239/239), and \emph{PIPEDA} (124/124/393/393). These counts pressure-test both coarse-grained patterns (privacy-related modules) and pinpoint evidence (specific sinks guarded by conditions), without biasing toward a single scope.

\paragraph{Why standard metrics are not enough (motivation for a new evaluation framework).}
Traditional per-metric views are indispensable for debugging behavior (e.g., MRR for early precision, nDCG for rank quality, Micro-F1 for label balance). In our setting, however, they \emph{fail to make stability visible}: a model can look strong on MAP at the file level yet collapse at the line level; it can show good Jaccard on one law’s taxonomy yet drift on another; and it can excel on Task~2’s labeling while struggling to \emph{find} the evidence in Task~1. In short, single metrics (and even per-task metric suites) do not surface the cross-granularity and cross-law \emph{robustness} practitioners need for credible compliance triage.

To address this, we introduce a \emph{stability-first evaluation framework} that aggregates standard metrics into interpretable composites, with explicit penalties for inconsistency:

\begin{itemize}
  \item \textbf{SGS} (Stability across Granularities for Task~1): for each retrieval metric, we combine performance at \emph{file}, \emph{module}, and \emph{line} using a variance-penalized harmonic mean, so the weakest scope meaningfully drags down the score. This reflects end-to-end localization: auditors must traverse these levels consistently.
  \item \textbf{RCS} (Regulation-wise Composite Score): within each law and task, we summarize multiple metrics using a TOPSIS-style aggregation with Mahalanobis distance \cite{hwang1981topsis,mahalanobis1936}. This attenuates double-counting correlated metrics (e.g., MRR vs.\ MAP) and yields a single, regulation-specific score per task.
  \item \textbf{CRGS} (Cross-Regulation Geometric Stability): we aggregate a model’s per-law composite for a task via a geometric mean with a variance penalty, rewarding methods that transfer across statutory taxonomies rather than spiking on only one.
  \item \textbf{OCS} (Overall Coupled Stability): we first \emph{couple} Task~1 and Task~2 within each law by a variance-penalized harmonic mean (localization \emph{and} judgment must both hold), and then aggregate across laws as in CRGS. Unless a model is balanced across tasks and jurisdictions, OCS will not inflate.
\end{itemize}

Unless stated otherwise, we use default penalty weights $(\alpha{=}1, \beta{=}2, \gamma{=}2, \delta{=}2)$ for granularity-, metric-, law-, and task-level coupling respectively (ablation in~\S\ref{sec:results}). The result is not a new single “leaderboard number” for its own sake, but a compact \emph{lens} that forces stability to surface where it matters operationally.

We evaluate six strong Large Language Models (LLMs)—\emph{gpt-4o}, \emph{claude-3.5-sonnet-20241022}, \emph{claude-3.7-sonnet-20250219}, \emph{qwen2.5-72b-instruct}, \emph{gemini-2.5-pro}, and \emph{o1}—under a unified, auditable setup. Three key insights emerge from our analysis. First, localization proves to be a significant bottleneck, with models like \emph{claude-3.5-sonnet-20241022} performing well in cross-granularity retrieval, while others, like \emph{gemini-2.5-pro}, struggle considerably. Second, fragment-level judgment shows stronger but uneven performance, with models excelling in specific jurisdictions but not consistently across all laws. Finally, when combining retrieval and judgment, the rankings shift, highlighting that strong judgment alone is insufficient if retrieval performance is unstable. This underscores the importance of evaluating models as holistic systems, balancing both tasks to ensure reliable performance across jurisdictions and statutory provisions.

\noindent
\textbf{Our contributions are shown as following:}
\begin{enumerate}
    \item We present a code-and-law benchmark that links Android code artifacts to native statutory articles in LGPD, PDPA, and PIPEDA, with reproducible anchors (commit/file/line).
    \item We introduce two tasks aligned with real-world compliance reviewer workflows—cross-granularity retrieval (Task~1) and fragment-level multi-label judgment (Task~2)—evaluated by standard metrics and a new \emph{stability-first} framework (SGS, RCS, CRGS, OCS) that highlights model robustness.
    \item We provide a comparative study of six LLMs under a unified setup, demonstrating that while models such as \emph{gpt-4o} excel in overall stability (OCS = 0.2736), they still exhibit uneven performance across tasks, jurisdictions, and granularities.
\end{enumerate}

\paragraph{Paper structure.}
Section~\S\ref{sec:background} presents background and the problem formulation; Section~\ref{sec:relatedwork} reviews related work in software engineering and legal-text analysis; Section~\ref{sec:benchmark-design} details the dataset and benchmark design; Section~\ref{sec:setup} outlines the experimental setup; Section~\ref{sec:results} presents the empirical results and analyses; Section~\ref{sec:discussion} discusses the implications; Section~\ref{sec:validity} reflects on threats to validity; and Section~\ref{sec:conclusion} concludes the paper.

We release the CompliBench dataset, prompts, and evaluation scripts at \url{https://github.com/Haoyi-Zhang/CompliBench} to encourage reproducible comparisons and advance regulation-aware software engineering.

\section{Background and Problem Formulation}
\label{sec:background}

\noindent
Compliance review for Android software straddles two worlds. On one side sits the concrete reality of repositories: multi-module builds, component lifecycles, manifest declarations, and SDK integrations that collectively determine when and how data is touched. On the other sits statute language that carves obligations into articles and sub-articles. A credible review must bridge these worlds in a way that is (i) \emph{locatable} in code and (ii) \emph{auditable} against the law as written. This section develops the technical and legal context for that bridge and then formalizes the two tasks our benchmark evaluates. We deliberately avoid the evaluation methodology here; details of metrics and aggregation live later alongside the dataset description to prevent redundancy.

\subsection{Android evidence: where obligations surface in code}
\label{subsec:android-evidence}

Android applications are decomposed into \textit{components} (activities, services, receivers, providers) that interact via intents and callbacks. Privacy-relevant behavior emerges from the interplay of three ingredients that scatter evidence across the codebase:

\begin{itemize}
  \item \textbf{Entry points and lifecycles.} Callback-driven execution (e.g., \textsf{onCreate}, \textsf{onResume}, foreground service starts) controls \emph{when} identifiers or telemetry may be read or transmitted.
  \item \textbf{Capabilities and declarations.} Manifests, permissions, Gradle settings, and library configuration expose \emph{what} the app is allowed or intends to do.
  \item \textbf{Data flows and sinks.} API calls, SDK hooks, and networking code encode \emph{how} data moves from sources to storage or external endpoints.
\end{itemize}

A large body of SE research targets precisely these facets. Static taint analyses such as FlowDroid and Amandroid model lifecycles and inter-component communication (ICC); IccTA captures cross-component leaks; dynamic approaches like TaintDroid track flows at runtime \cite{arzt2014flowdroid,wei2014amandroid,li2015iccta,enck2010taintdroid}. Ecosystem resources like AndroZoo enable scale studies and robust sampling \cite{allix2016androzoo}. Beyond these, work on permission modeling and intent resolution---PScout, EPICC, and whole-program information-flow analysis in DroidSafe---has clarified how capabilities are declared and how ICC actually resolves in practice \cite{au2012pscout,octeau2013epicc,gordon2015droidsafe}. This literature provides \emph{explainable} code evidence (e.g., a source--sink path or an ICC map), which is invaluable—but by itself it does not decide \emph{which statutory article} an engineer must satisfy. That final step necessarily engages the statute.

\subsection{Reviewer workflow and traceability requirements}
\label{subsec:workflow-traceability}

In an engineering organization, reviewers typically operate in three passes, each with its own evidence burden. First, they \emph{scan} the repository to flag suspicious modules (e.g., analytics, telemetry, identifier management). Second, they \emph{drill down} within flagged areas to \emph{lines} that perform data collection, transformation, or transmission. Third, they \emph{justify} the finding against a statute by citing specific article identifiers. The first two passes resemble repository-scale retrieval; the last requires statute-aware judgment. Crucially, any proposed issue must be \emph{traceable}: a concrete file path and line anchor on the code side, and a concrete article number on the law side. These operational constraints motivate our decision to keep immutable provenance (commit, path) and to retain native article numbering in the label space.

\subsection{From statute text to code obligations}
\label{subsec:statute-to-code}

Data-protection laws articulate duties as numbered \emph{articles}. Although themes recur across jurisdictions (consent, purpose specification, notification, retention, safeguards), their taxonomies and phrasing diverge.\footnote{For example, the scope and exceptions of "consent" differ across LGPD, PDPA, and PIPEDA.} Prior work in legal/policy NLP has demonstrated that textual artifacts---privacy policies in particular---can be analyzed and classified at scale \cite{wilson2016opp115,harkous2018polisis}. This broader theme of analyzing software-related artifacts to identify issues is well-established in software engineering, with researchers mining other forms of text, such as app reviews to find bugs \cite{tang2024app}, or analyzing code artifacts like commits to detect silent security patches, sometimes leveraging large language models (LLMs) for augmentation \cite{10.1145/3749370}. However, within the specific domain of legal compliance, much of the empirical attention on textual analysis has centered on GDPR \cite{gdpr_eu}.

In contrast, there are few \emph{code-grounded}, article-specific benchmarks for other national regimes. Our benchmark targets three statutes that, to our knowledge, lack such resources: Brazil's LGPD, Singapore's PDPA, and Canada's PIPEDA \cite{lgpd_govbr,pdpa_agc,pipeda_justice}. Two design choices follow. First, we preserve \emph{native article numbering} and identifiers, avoiding remapping into bespoke label sets that could obscure legal nuance. Second, every benchmark instance is \emph{scoped to a single law} to prevent inadvertent cross-law conflation while still enabling comparative analysis later.

\subsection{Granularity, ambiguity, and typical failure modes}
\label{subsec:granularity-failures}

Locating and judging potential violations is complicated by (at least) three ambiguity sources:

\begin{enumerate}
  \item \textbf{Granularity drift.} A model that is confident at \textsc{file}-level (e.g., \texttt{analytics/}) may fail at \textsc{line}-level because multiple call sites share nearly identical API shapes; only specific guards or parameterizations trigger an obligation.
  \item \textbf{ICC and lifecycle effects.} Android intent resolution and callback ordering alter whether a candidate sink is reachable in the user-visible context \cite{octeau2013epicc}. Lifecycle-aware static analyses exist, but their abstractions can misalign with snippet-limited reasoning and vice versa \cite{gordon2015droidsafe,arzt2014flowdroid}.
  \item \textbf{Label long tails.} Statutes include infrequent but critical articles. Systems that overfit frequent obligations (e.g., generic consent) risk missing specific duties (e.g., retention grounds or notification contents), which undermines auditability.
\end{enumerate}

These challenges inform both our corpus design (Section~\S\ref{sec:benchmark-design}) and the way we pose tasks to models (below).

\subsection{Corpus and instance construction (high-level principles)}
\label{subsec:corpus-principles}

This work evaluates systems on \emph{real} Android repositories (Java/Kotlin/XML) rather than synthetic snippets. At a high level:

\begin{itemize}
  \item \textbf{Immutable provenance.} Each instance binds to a commit hash and a stable file path to ensure that evidence is reproducible.
  \item \textbf{Granularity spectrum.} Evidence is anchored at three code granularities—\textsc{file}, \textsc{module}, \textsc{line}—plus a \textsc{snippet} view for contextual judgment. This mirrors how human reviewers drill down from broad suspicion to exact locations.
  \item \textbf{Law-native labels.} Ground-truth labels are sets of article identifiers from exactly one statute (LGPD, PDPA, or PIPEDA), recorded as they appear in the statute text.
\end{itemize}

These principles are motivated by practice rather than convenience: a compliance engineer must both \emph{find} the code and \emph{defend} the mapping to the statute, which is why we retain anchors and article numbers verbatim. Detailed dataset statistics and construction steps appear in Section~\S\ref{sec:benchmark-design}; we only record the rationale here to keep the narrative coherent.

\subsection{Problem formulation}
\label{subsec:problem-formulation}

We formalize two tasks that together approximate a realistic compliance-review workflow. Let $\mathcal{R}$ denote the set of repositories under consideration. For each law $L \in \{\textsc{LGPD}, \textsc{PDPA}, \textsc{PIPEDA}\}$, let $\mathcal{A}^{(L)}$ be its article set; the global label space is the disjoint union $\mathcal{A} \coloneqq \bigsqcup_{L} \mathcal{A}^{(L)}$. Let $\mathcal{G}\coloneqq\{\textsc{file},\textsc{module},\textsc{line}\}$ be the set of code granularities, and let $\mathcal{S}$ be the set of curated snippet windows.

\paragraph{Task~1: Cross-granularity localization (retrieval).}
Given a repository $r\in\mathcal{R}$, a target law $L$, and a granularity $g\in\mathcal{G}$, the system must rank candidate artifacts $\mathcal{C}_{r,g}$ to surface those associated with a potential violation grounded in $L$. Formally, the model parameters $\theta$ induce a scoring function
\[
  f_{\theta} : \mathcal{C}_{r,g} \times \Phi \to \mathbb{R},
\]
where $\Phi$ denotes optional side information (e.g., neighboring lines, manifest fragments). Ground truth is a non-empty relevance set $\mathcal{Y}\subseteq \mathcal{C}_{r,g}$. The output is a total preorder over $\mathcal{C}_{r,g}$.

\textit{Why this is hard.} Localization must be \emph{consistent} across $g\in\{\textsc{file},\textsc{module},\textsc{line}\}$. A system that proposes the right file but consistently misses the decisive line anchor is of limited utility for remediation. Moreover, repositories can contain multiple candidate sites with similar API shapes; discriminating the one that actually triggers an article obligation is non-trivial (e.g., guarded vs.\ unguarded data transmission).

\paragraph{Task~2: Fragment-level statutory judgment (multi-label).}
Given a curated snippet $s\in\mathcal{S}$ extracted from a repository and a law $L$, the system must predict the full set of applicable articles $\mathcal{Y}\subseteq \mathcal{A}^{(L)}$. We view this as a set prediction problem with a hypothesis $h_{\theta}: \mathcal{S} \rightarrow 2^{\mathcal{A}^{(L)}}$.

\textit{Why this is hard.} Multiple obligations often co-occur (e.g., consent \emph{and} purpose specification). Articles also vary in granularity: some are general principles; others enumerate precise requirements. Finally, class frequency is long-tailed. Preserving \emph{native} identifiers prevents collapsing distinct obligations and supports downstream legal review when a model’s rationale must be audited against statute text.

\subsection{Inputs, outputs, and constraints}
\label{subsec:io-constraints}

We summarize the machine-facing interfaces without prescribing the scoring protocol (which appears later to avoid duplication).

\smallskip
\noindent\textbf{Inputs.}
Each instance provides: (i) a repository pointer $(r,\text{commit},\text{path})$; (ii) a target law $L$; and either (iii-a) a granularity tag $g\in\mathcal{G}$ with implicit candidate set $\mathcal{C}_{r,g}$ (for Task~1), or (iii-b) a curated snippet $s$ with minimal context (for Task~2).

\smallskip
\noindent\textbf{Outputs.}
For Task~1, a ranked list over $\mathcal{C}_{r,g}$. For Task~2, a predicted article set $\widehat{\mathcal{Y}}\subseteq \mathcal{A}^{(L)}$.

\smallskip
\noindent\textbf{Constraints and non-goals.}
(1) Instances are evaluated against exactly one statute to avoid cross-law conflation.  
(2) We do not evaluate runtime network traces or policy text; those are complementary artifacts with different collection trade-offs \cite{wilson2016opp115,harkous2018polisis}.  
(3) Repository state is immutable (commit-pinned) for replicability.  
(4) Snippet windows include only the minimum surrounding lines required for an informed decision; entire files are avoided to reduce confounds unrelated to reasoning.

\subsection{Research questions}
\label{subsec:rqs}

To connect with the introduction and guide the empirical study that follows, we frame four questions that our formulation enables:

\begin{itemize}
  \item \textbf{RQ1 (Localization).} How reliably can modern LLM-based systems localize potential violations across \textsc{file}, \textsc{module}, and \textsc{line} granularities in real repositories?
  \item \textbf{RQ2 (Judgment).} Given a localized code context, to what extent can these systems assign the complete set of applicable statutory articles, including infrequent obligations?
  \item \textbf{RQ3 (Generalization).} Do observed behaviors transfer across LGPD, PDPA, and PIPEDA, or are successes statute-specific?
  \item \textbf{RQ4 (Coupling).} When localization and judgment are considered together at the statute level, does the system remain useful for end-to-end compliance triage?
\end{itemize}

These questions structure the remainder of the paper. Section~\S\ref{sec:relatedwork} situates our formulation within SE program analysis and legal/policy NLP. Section~\S\ref{sec:benchmark-design} then specifies the dataset and evaluation protocol used to answer the RQs, followed by setup and results.

\section{Related Work}
\label{sec:relatedwork}

\noindent
We review four strands that intersect in \textit{CompliBench}: (i) Android privacy analyses and ICC mapping that uncover \emph{where/how} data moves in code; (ii) resources and studies around mobile data practices and SDK behavior; (iii) legal/policy NLP that reasons over \emph{textual} obligations; and (iv) code understanding, retrieval, and repository-level LLM benchmarks. Our goal is to situate a benchmark that \emph{links} concrete Android evidence (file$\rightarrow$module$\rightarrow$line) to \emph{native} statutory articles (LGPD/PDPA/PIPEDA), and that evaluates both localization and statute-aware judgment under a stability-first philosophy.

\subsection{Android privacy analyses and inter-component communication}
\label{subsec:rw-android-analysis}

Static taint analysis has been the workhorse for pinpointing sensitive flows. \textsc{FlowDroid} introduced lifecycle-, context-, and object-sensitive analysis for Android, enabling precise source–sink tracing within components \cite{arzt2014flowdroid}; \textsc{Amandroid} generalized to inter-component data flow \cite{wei2014amandroid}; and \textsc{IccTA} operationalized leak detection across components in real apps \cite{li2015iccta}. Dynamic tracking complements static summaries: \textsc{TaintDroid} demonstrated real-time, system-wide tainting on commodity devices \cite{enck2010taintdroid}. These techniques seeded widely used building blocks: permission models (\textsc{PScout}) \cite{au2012pscout}, ICC resolvers (EPICC) \cite{octeau2013epicc}, and whole-program information-flow analysis (\textsc{DroidSafe}) \cite{gordon2015droidsafe}. Benchmarks such as \textsc{DroidBench} and automatic source/sink discovery via \textsc{SuSi} strengthened evaluation coverage and reduced manual labeling costs \cite{arzt2014soap,arzt2014susi}.

\emph{Relation to our work.} These systems establish \emph{where/how} personal data can flow and have shaped our instance construction (commit-pinned paths; file/line anchors; ICC-relevant context). However, they do not answer \emph{which statute article} a given code path may implicate, nor do they assess cross-granularity \emph{stability} in localization when moved from file$\rightarrow$module$\rightarrow$line. \textit{CompliBench} complements them by tying localization outputs to \emph{law-native} labels and by explicitly testing consistency across granularities and statutes.

\subsection{Mobile data practices, identifiers, and SDKs}
\label{subsec:rw-mobile-ecosystem}

A parallel line of empirical work studies real-world collection and sharing of personal data in mobile apps. Large-scale corpus efforts (e.g., \textsc{AndroZoo}) enabled ecosystem measurements at scale \cite{allix2016androzoo}. Closer to on-device behavior, Reardon et~al.\ exposed how apps exfiltrate or circumvent protections around identifiers and telemetry, highlighting the prevalence of background collection and third-party SDK mediation \cite{reardon2019fifty}. These findings underscore why compliance reviews must not stop at “any use of a sink” but instead pinpoint \emph{which flows, under which conditions}, rise to statutory obligations. Our corpus reflects that reality by including modules and lines that are superficially similar at the API level but differ in guards, destinations, or data semantics—the very details auditors must cite.

\subsection{Legal/policy NLP and statutory reasoning}
\label{subsec:rw-legal-nlp}

Policy-focused NLP has shown that privacy commitments in text can be parsed and queried at scale: OPP-115 provides annotated policy segments, and \textsc{Polisis} automates multilabel policy analysis \cite{wilson2016opp115,harkous2018polisis}. PrivacyQA casts policy understanding as question answering \cite{ravichander2019privacyqa}; legal-domain pretraining and evaluation benchmarks (LEGAL-BERT, LEXGLUE) broaden task coverage beyond privacy to charge prediction, statute entailment, and case classification \cite{chalkidis2020legalbert,chalkidis2021lexglue}. Much of this activity orbits GDPR on the text side. 

\emph{Relation to our work.} We invert the usual vantage point: rather than analyzing \emph{policies}, we evaluate whether systems can map \emph{code evidence} to \emph{statute articles}. Maintaining law-native article identifiers (LGPD/PDPA/PIPEDA) preserves auditability and avoids conflating obligations through bespoke taxonomies. This makes our benchmark complementary to policy NLP and closer to the evidence chain a compliance engineer must produce.

\subsection{Code search, representation learning, and LLM benchmarks}
\label{subsec:rw-code-llm}

Early code search relied on IR measures and textual heuristics; large pretraining has since improved retrieval and summarization. \textsc{CodeSearchNet} established shared tasks for semantic code search \cite{husain2019codesearchnet}. Pretrained models such as \textsc{CodeBERT}, \textsc{GraphCodeBERT}, and \textsc{CodeT5} learn joint code–NL representations (with data-flow or identifier-aware objectives), improving code search, summarization, and generation \cite{feng2020codebert,guo2021graphcodebert,wang2021codet5}. Broader surveys synthesize the trajectory of learning-from-code \cite{allamanis2018survey}. In parallel, SE LLM benchmarks track step-changes from synthesis to repository reasoning: HumanEval/MBPP (function-level synthesis), CodeXGLUE (multi-task code understanding), and SWE-bench (repo-level issue resolution) \cite{chen2021humaneval,austin2021mbpp,lu2021codexglue,jimenez2023swe-bench}. Newer unified encoders like \textsc{UniXcoder} target cross-modal/code tasks with improved retrieval and generation capabilities \cite{guo2022unixcoder}.

\emph{Relation to our work.} These datasets and models are essential for code intelligence writ large, but they rarely demand \emph{statute-grounded} judgments nor place pressure on multi-granularity localization in \emph{Android} repos. \textit{CompliBench} complements repository-level tasks by coupling (i) retrieval across file$\rightarrow$module$\rightarrow$line with (ii) multilabel article assignment, and by reporting results per statute with law-native identifiers to support external audit.

\subsection{Positioning}
\label{subsec:rw-positioning}

Across these strands, a consistent gap emerges: existing tools tell us \emph{where} sensitive data may flow or how to retrieve a relevant snippet; policy NLP tells us \emph{what} an organization claims or what a law entails. What compliance engineers need in practice is the \emph{joint} capability: \emph{find} the right locations in a real repository and \emph{justify} them against concrete article numbers in the governing statute. \textit{CompliBench} is designed precisely for that joint evaluation, and the remainder of this paper shows—empirically—that strong models can excel on one side of the bridge while faltering on the other, motivating our stability-first evaluation lens.

\section{Benchmark Design}
\label{sec:benchmark-design}

\begin{figure*}[t]
  \centering
  \includegraphics[width=\textwidth]{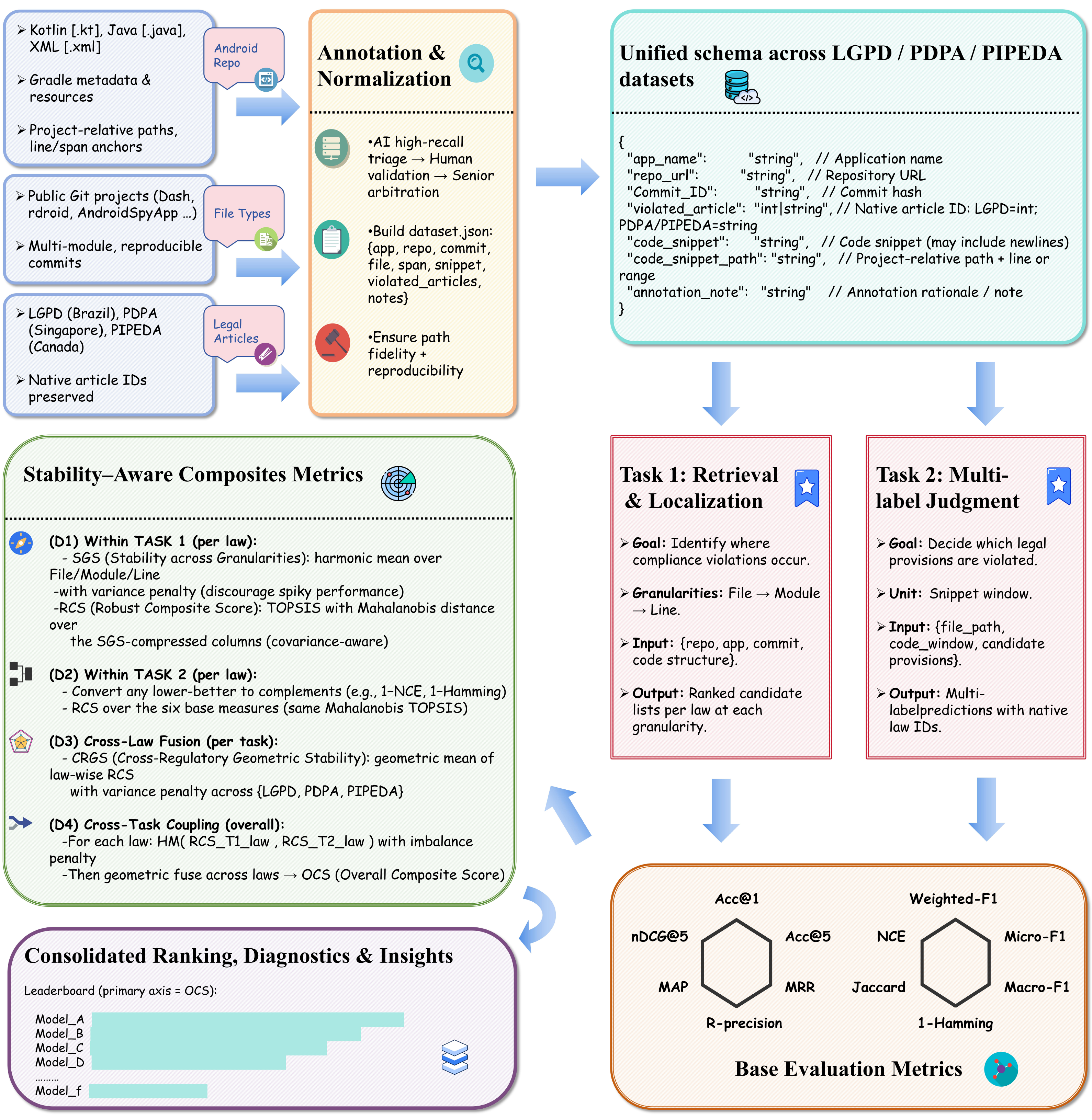}
  \caption{\textbf{CompliBench overview.} Android repositories and statutory articles are transformed into two evaluation tracks: \emph{Task~1} (retrieval at file/module/line granularity) and \emph{Task~2} (multi-label classification at the snippet level). Standard metrics feed into stability-aware composites (SGS, RCS, CRGS, OCS), yielding cross-granularity and cross-jurisdiction insights. The figure adopts a staged layout (\emph{Repos} $\rightarrow$ \emph{Annotation \& Normalization} $\rightarrow$ \emph{Schema \& Tasks} $\rightarrow$ \emph{Metrics} $\rightarrow$ \emph{Insights}) with branching evaluation tracks and consolidated ranking diagnostics.}
  \label{fig:overview}
\end{figure*}

\noindent
This section presents the design of \emph{CompliBench}, a benchmark that evaluates whether large language models (LLMs) can detect real-world compliance violations in Android software. As illustrated in Figure~\ref{fig:overview}, CompliBench organizes the end-to-end pipeline into five interconnected stages. 

First, we begin from \emph{real Android repositories} (multi-module Kotlin/Java/XML projects) combined with statutory corpora spanning three jurisdictions (LGPD, PDPA, PIPEDA). Second, through annotation and normalization, raw code and legal evidence are consolidated into structured, machine-usable artifacts with commit-level reproducibility. Third, the pipeline bifurcates into two complementary tasks: \textbf{Task~1} (\emph{retrieval \& localization}) tests whether a model can surface implicated artifacts at the \emph{file}, \emph{module}, and \emph{line} levels, while \textbf{Task~2} (\emph{multi-label judgment}) probes whether a model can map curated code snippets to the correct set of violated statutory articles. Fourth, both tasks are evaluated with established metrics from information retrieval and multi-label learning, including Acc@k, R-Precision, MRR, MAP, nDCG, Micro/Macro-F1, Weighted-F1, Jaccard, Hamming loss, and normalized coverage error. Finally, CompliBench integrates these base metrics into \emph{stability-aware composite scores} (SGS, RCS, CRGS, and OCS), enabling principled model comparison across granularities and jurisdictions.

\noindent
Figure~\ref{fig:overview} further emphasizes the benchmark’s dual focus on \emph{diagnostic transparency} and \emph{system-level comparability}. By structuring both tasks under a unified schema with immutable provenance, the benchmark ensures that individual predictions can be traced back to concrete code spans and article references, while aggregated composites mitigate spiky or law-specific performance. The inclusion of side-panel diagnostics and leaderboard views supports not only ranking but also error analysis, highlighting when models succeed at broad recall yet fail at fine-grained localization, or when retrieval and judgment performance diverge. In this way, the figure encapsulates both the micro-level auditability of model outputs and the macro-level robustness of cross-jurisdictional assessment.

\noindent
This modular design allows CompliBench to serve as both a research benchmark and a diagnostic tool, bridging practical auditing workflows with reproducible experimental protocols. The remainder of this section elaborates its constituent elements: Section~\ref{sec:design-principles} introduces the guiding design principles, Section~\ref{sec:dataset} details dataset construction, Section~\ref{sec:task-design} formalizes the two tasks, and Section~\ref{sec:metrics} specifies the evaluation metrics, including the stability-aware composites.

\subsection{Design Principles}
\label{sec:design-principles}

\noindent
Guided by the end-to-end workflow in Fig.~\ref{fig:overview}, we distill eight principles that operationalize \emph{CompliBench} as a realistic, auditable, and cross-jurisdictional benchmark.

\begin{principlebox}[P1. Realism-first, code-grounded evaluation]
Compliance in mobile software emerges from concrete code patterns, control and data flows, and configuration files. CompliBench therefore anchors all instances in \emph{real Android repositories} (multi-module projects with Kotlin, Java, and XML), preserving file structure and contextual lines to make the tasks faithful to developer-facing scenarios and auditable by reviewers (\S\ref{sec:dataset}). See also empirical analyses of Android privacy flows and inter-component communication that motivate code-grounded evidence \cite{arzt2014flowdroid, Wei2012EPICC, enck2010taintdroid}.
\end{principlebox}

\begin{principlebox}[P2. Dual-track probing of complementary abilities]
Detecting violations entails both \emph{finding} where a violation may occur and \emph{judging} which provisions are implicated. We operationalize these abilities via two tracks: \textbf{Task~1} frames compliance detection as retrieval and localization across file, module, and line granularities, while \textbf{Task~2} frames it as multi-label classification for a snippet. This separation reduces construct underrepresentation and allows targeted error analysis (\S\ref{sec:task-design}). This mirrors established distinctions between retrieval and ranking versus multi-label judgment in information retrieval and machine learning \citep{manning2008ir, zhang2014multilabel}.
\end{principlebox}

\begin{principlebox}[P3. Multi-granularity supervision for practical utility]
Audits rarely end at a single granularity. By supervising file, module, and line in Task~1, the benchmark captures progressively finer localization fidelity, enabling practitioners to trade off triage accuracy against investigation cost. Granularity-consistent scoring also reveals whether a model’s gains are stable or brittle across levels (\S\ref{sec:task-design}, \S\ref{sec:metrics}). Granularity interacts with ranking depth, as classical IR metrics emphasize early precision \citep{jarvelin2002cumulated, manning2008ir}.
\end{principlebox}

\begin{principlebox}[P4. Cross-jurisdiction coverage for external validity]
The same code behaviors can map to different legal obligations across jurisdictions. We include three widely adopted privacy regimes—LGPD \citep{lgpd_govbr}, PDPA \citep{pdpa_agc}, and PIPEDA \citep{pipeda_justice}—to evaluate whether models generalize beyond a single article taxonomy and to quantify jurisdictional variance as a first-class signal (\S\ref{sec:dataset}).
\end{principlebox}

\begin{principlebox}[P5. Standard metrics: stability-aware aggregation thereafter]
Per-task metrics follow established practice: Acc@k, R-Precision, MRR, MAP, and nDCG for retrieval and ranking, and micro, macro, and weighted F1, Jaccard, Hamming loss, and normalized coverage error for multi-label classification. To avoid misleading conclusions from any single metric, CompliBench introduces \textbf{SGS} (a cross-granularity stability score), \textbf{RCS} (a TOPSIS-based composite with Mahalanobis distance to respect inter-metric covariance), and \textbf{CRGS} (a cross-regulation geometric score with variance penalties) to reward models that are not only strong but also \emph{stable} across metrics, granularities, and jurisdictions (\S\ref{sec:metrics}). Base measures follow standard IR and multi-label practice \citep{manning2008ir, jarvelin2002cumulated, zhang2014multilabel}, while our composites explicitly model inter-metric dependence (Mahalanobis) and distance-to-ideal ranking (TOPSIS) \citep{mahalanobis1936, hwang1981topsis}.
\end{principlebox}

\begin{principlebox}[P6. Transparent key matching and diagnosability]
For localization, imperfect key alignment between predictions and gold annotations (for example due to path normalization or snippet offsets) can confound measurement. We adopt explicit matching policies (strict versus relaxed), report coverage statistics (gold versus matched items), and expose per-level diagnostics to make ceiling effects and error modes observable rather than implicit (\S\ref{sec:metrics}). Disclosed matching policies are consistent with best practice in retrieval evaluation and error analysis \citep{manning2008ir}.
\end{principlebox}

\begin{principlebox}[P7. Reproducibility and auditability by design]
All scripts expose configuration knobs for normalization, key-matching policy, and composite weighting; outputs record settings alongside results. Datasets preserve code context and article references to facilitate external audit, while evaluation tables and machine-readable dumps enable independent recomputation and meta-analysis (\S\ref{sec:dataset}, \S\ref{sec:metrics}). Reporting mirrors conventional IR and ML result disclosures to facilitate independent recomputation \citep{manning2008ir, zhang2014multilabel}.
\end{principlebox}

\begin{principlebox}[P8. Cost and latency awareness while minimizing provider bias]
The inference harness abstracts over model providers and logs throughput and latency, allowing fair comparisons of systems-level trade-offs while keeping provider-specific knobs (temperature, context length) explicit in the setup section. This discourages overfitting conclusions to any single API configuration and supports deployment-minded interpretation. Operational details are specified in \S\ref{sec:setup}.
\end{principlebox}

\subsection{Dataset Construction}
\label{sec:dataset}

\noindent
This subsection explains how the \emph{raw} CompliBench corpus is assembled, normalized, and documented \emph{prior} to any task-specific materialization. We focus on (i) \emph{provenance \& sampling}, (ii) a \emph{human–AI} curation loop that balances coverage with fidelity, and (iii) released artifacts that enable independent audit and replication. The formulation of Task~1/Task~2 is deferred to Section~\ref{sec:task-design} to maintain a clean separation between data construction and evaluation design.

\paragraph{Provenance and sampling.}
We ground the corpus in public Android repositories that (a) build or can be parsed without proprietary SDKs, (b) include Kotlin/Java sources with XML resources and Gradle metadata, and (c) exhibit privacy-relevant behaviors (e.g., identifier/sensor collection, inter-process/background services, remote transmission, on-device storage). To support cross-regulation analysis, we curate \emph{native} identifiers for three regimes—Brazil’s \mbox{LGPD} \citep{lgpd_govbr}, Singapore’s \mbox{PDPA} \citep{pdpa_agc}, and Canada’s \mbox{PIPEDA} \citep{pipeda_justice}—and preserve their numbering and citation styles (e.g., ``Art.~7'', ``s.~13'', ``\S\,4.3'') verbatim. Every instance records its repository URL and commit hash, making the corpus re-extractable at an immutable snapshot (and auditable down to file/line anchors).

\paragraph{Human–AI coordinated pipeline.}
First, we parse repositories into snippets with stable pointers (\texttt{file:line} or short spans) and apply a high-recall AI triage that proposes \emph{candidate} provisions per snippet and jurisdiction. Second, compliance-aware annotators verify, correct, or reject these candidates against statutory text and code context; disagreements are arbitrated by a senior reviewer. The result is a \emph{raw}, expert-validated corpus that links code evidence to native article IDs while retaining reviewer notes. These notes are released for transparency yet held out from model inputs to avoid rationale leakage during evaluation.

\paragraph{Released schema.}
The raw artifact (\texttt{dataset.json}) exposes (i) application- and repository-level provenance, (ii) immutable code pointers and excerpts, (iii) jurisdiction-native labels, and (iv) expert rationale. This enables bidirectional audit: from any article and snippet back to the exact commit and lines, and from a repository snapshot forward to all curated snippets it contains.

\begin{table}[H]
  \centering
  \caption{Schema of the raw dataset (\texttt{dataset.json}).}
  \label{tab:raw-schema}
  \begin{tabular}{@{}lll@{}}
    \toprule
    \textbf{Field} & \textbf{Type} & \textbf{Description} \\
    \midrule
    app\_name   & String  & Application name \\
    repo\_url   & String  & Repository URL \\
    commit\_id  & String  & Commit hash for snapshot \\
    article\_id & Int/List & Linked legal article(s) \\
    file\_path  & String  & Path and line range \\
    snippet     & String  & Extracted code snippet \\
    note        & String  & Expert annotation \\
    \bottomrule
  \end{tabular}
\end{table}

\noindent\textit{Discussion.}
Separating \texttt{file\_path} from \texttt{snippet} allows strict pointer checks (path normalization, span validity) without conflating location with content, while leaving \texttt{article\_id} native to each regulation avoids lossy remapping at construction time. Expert \texttt{note} fields rationalize code–law links for human readers yet remain excluded from prompts, keeping subsequent evaluations code-grounded.

\paragraph{Corpus characteristics across jurisdictions and granularities.}
We report counts at four granularities: \emph{file}, \emph{module} (file-scoped unit), \emph{line} (or short span), and \emph{snippet}. Although overall volumes differ by regime, the profiles are intentionally heterogeneous: PIPEDA offers denser line/snippet coverage (facilitating fine-grained localization), whereas LGPD and PDPA appear more balanced between file and module, aiding component-level audits.

\begin{table}[H]
  \centering
  \caption{Dataset statistics by jurisdiction and granularity.}
  \label{tab:granularity}
  \begin{tabular}{@{}lrrrr@{}}
    \toprule
    \textbf{Law} & \textbf{File} & \textbf{Module} & \textbf{Line} & \textbf{Snippet} \\
    \midrule
    LGPD   & 90  & 90  & 208 & 208 \\
    PDPA   & 77  & 78  & 239 & 239 \\
    PIPEDA & 124 & 124 & 393 & 393 \\
    \bottomrule
  \end{tabular}
\end{table}

\noindent\textit{Discussion.}
The near equality of \emph{file} and \emph{module} counts suggests tight coupling between source files and their primary units (e.g., classes/components) in our selection; this simplifies reproduction and review. Higher \emph{line}/\emph{snippet} tallies reflect many-to-one expansion from files to granular violation anchors.

\paragraph{Repository coverage and diversity.}
To avoid dataset dominance by a few projects, we cross-index jurisdictions against contributing repositories and observe broad spread (e.g., \emph{Dash}, \emph{rdroid}, \emph{AndroidSpyApp}). PIPEDA shows the largest absolute counts in several codebases, driven by denser snippet extraction; LGPD and PDPA show complementary coverage patterns (e.g., \emph{L3MON} for LGPD; \emph{rdroid}/\emph{Dash} for both). Since commit hashes are preserved, these distributions are fully auditable.

\begin{table}[H]
  \centering
  \caption{Repository coverage by jurisdiction (number of raw instances).}
  \label{tab:repo-matrix}
  \begin{tabular}{@{}lrrrrrrrr@{}}
    \toprule
    \textbf{Law} &
    \makecell{pounce-\\keys} &
    AIRAVAT &
    \makecell{Android\\SpyApp} &
    L3MON &
    rdroid &
    Dash &
    \makecell{Privacy\\Breacher} &
    \makecell{Rafel\\Rat} \\
    \midrule
    LGPD   & 26 & 0  & 81 & 21 & 47  & 138 & 15 & 75 \\
    PDPA   & 38 & 0  & 61 & 0  & 77  & 89  & 23 & 82 \\
    PIPEDA & 42 & 29 & 94 & 23 & 121 & 192 & 28 & 102 \\
    \bottomrule
  \end{tabular}
\end{table}

\noindent\textit{Discussion.}
Overlaps (repositories present across laws) enable controlled cross-law comparisons on similar technical artifacts, whereas asymmetries (law-specific repositories) naturally induce out-of-domain conditions—useful later for transfer analyses but, at this stage, simply a property of the corpus we document.

\paragraph{Cross-law thematic alignment.}
Although article numbering differs by regime, many obligations align thematically (consent, notice, safeguards, transfers). We therefore provide a theme-centric \emph{alignment seed} at construction time. This is not a doctrinal equivalence claim; it is a pragmatic bridge for aggregate reporting and later ablations that \emph{preserves} native identifiers.

\begin{table}[H]
  \centering
  \caption{Cross-law mapping of compliance themes.}
  \label{tab:alignment}
  \begin{tabular}{@{}lrrrrrr@{}}
    \toprule
    \textbf{Law} & Consent & Notice & Collection & Retention & Security & Transfer \\
    \midrule
    LGPD   & 7   & 6   & 6   & 15  & 46  & 33 \\
    PDPA   & 13  & 20  & 18  & 25  & 24  & 26 \\
    PIPEDA & 4.3 & 4.2 & 4.4 & 4.5 & 4.7 & 4.1 \\
    \bottomrule
  \end{tabular}
\end{table}

\noindent\textit{Discussion.}
For example, \emph{Consent} aligns LGPD Art.~7, PDPA s.~13, and PIPEDA \S\,4.3; \emph{Security} aligns LGPD Art.~46, PDPA s.~24, and PIPEDA \S\,4.7 \citep{lgpd_govbr,pdpa_agc,pipeda_justice}. These anchors enable apples-to-apples summaries while keeping traceability to statutory text.

\paragraph{Pairwise overlap of aligned provisions.}
To quantify comparability, we compute a pairwise overlap matrix over the thematic anchors. LGPD–PDPA and PDPA–PIPEDA exhibit higher overlaps than LGPD–PIPEDA, indicating that certain obligation families admit stronger cross-law bridges. We emphasize that this is a \emph{dataset property} reported for transparency; task evaluation remains strictly native-ID.

\begin{table}[H]
  \centering
  \caption{Overlap matrix of compliance themes (darker = more overlap).}
  \label{tab:overlap-heat}
  \begin{tabular}{@{}lccc@{}}
    \toprule
    & LGPD & PDPA & PIPEDA \\
    \midrule
    LGPD   & --- & \cellcolor{blue!40}18 & \cellcolor{blue!20}12 \\
    PDPA   & \cellcolor{blue!40}18 & --- & \cellcolor{blue!30}15 \\
    PIPEDA & \cellcolor{blue!20}12 & \cellcolor{blue!30}15 & --- \\
    \bottomrule
  \end{tabular}
\end{table}

\noindent\textit{Discussion.}
Reporting overlap at the \emph{theme} level (rather than raw article counts) avoids conflating editorial granularity with doctrinal scope, yielding a more meaningful view of where cross-regime comparisons are feasible.

\paragraph{Label frequency and long-tail structure.}
The ten most frequent native identifiers per jurisdiction reveal heavy tails: common obligations (e.g., consent, notice, safeguards) recur across repositories, while rarer provisions (special categories, certain cross-border conditions) appear sparsely. We preserve this reality without rebalancing, leaving class-imbalance handling to the evaluation design in Section~\ref{sec:task-design}.

\begin{table}[H]
  \centering
  \caption{Top-10 frequent article labels by jurisdiction.}
  \label{tab:top10}
  \begin{tabular}{@{}lrrrrrrrrrr@{}}
    \toprule
    \textbf{Law} & \textbf{\#1} & \textbf{\#2} & \textbf{\#3} & \textbf{\#4} & \textbf{\#5} &
    \textbf{\#6} & \textbf{\#7} & \textbf{\#8} & \textbf{\#9} & \textbf{\#10} \\
    \midrule
    LGPD   & 7(90) & 12(88) & 5(72) & 6(64) & 33(24) & 46(15) & 15(14) & 11(13) & 8(12) & 34(11) \\
    PDPA   & 24(123) & 13(38) & 20(36) & 25(34) & 26(30) & 14(28) & 18(22) & 28(22) & 27(11) & 15(10) \\
    PIPEDA & 4.3(166) & 4.7(157) & 4.4(99) & 4.2(71) & 4.5(62) & 4.8(45) & 4.1(16) & 4.9(11) & 4.6(4) & --- \\
    \bottomrule
  \end{tabular}
\end{table}

\paragraph{Summary.}
In sum, the construction process yields a code-grounded, expert-validated corpus with (i) strong provenance guarantees, (ii) multi-granularity anchors, (iii) broad repository coverage, and (iv) principled cross-law comparability via theme-level alignment and quantified overlaps. These are \emph{dataset-level} properties that precede task shaping; Section~\ref{sec:task-design} builds on this foundation to define model-facing inputs/outputs and evaluation views.

\subsection{Task Design} 
\label{sec:task-design}

\noindent
Building on the expert-validated raw corpus in Section~\ref{sec:dataset}, \emph{CompliBench} defines two complementary evaluation views that reflect how human auditors reason about compliance in Android applications: (i) \emph{retrieval and localization} across multiple code granularities, and (ii) \emph{multi-label judgment} for a concrete snippet. The entire corpus serves as a \emph{gold standard}: no training, development, or test splits are created, and systems are not adapted on any portion of CompliBench. This design choice emphasizes intrinsic model capabilities under a fixed, auditable dataset, simplifying reproducibility across jurisdictions.

\paragraph{Design objectives and constraints.}
We identify four core objectives driving the task formulation:
\begin{itemize}[leftmargin=1.35em, itemsep=0.28em, topsep=0.1em]
  \item \textbf{O1: Separate \emph{where} from \emph{what}.} Audits follow a two-step process: first, investigators locate potentially problematic code, then determine the implicated provisions. Combining both steps into a single prediction obscures error sources. Therefore, we design two independent views that evaluate these stages separately.
  \item \textbf{O2: Preserve code and law \emph{as-is}.} The native file structure, line anchors, and jurisdiction-specific article identifiers are retained. This avoids lossy remapping at task time and supports subsequent aggregation by theme without altering the gold labels.
  \item \textbf{O3: Be granular yet stable.} Localization must scale from coarse (\emph{file}) to fine (\emph{line or short span}) without introducing brittle anchors. We apply deterministic normalization and discard pointers that cannot be re-validated at the referenced commit.
  \item \textbf{O4: Prevent evaluation leakage.} While free-text expert notes provide valuable insights for human audit, they may inadvertently leak answers. These notes are preserved for transparency but \emph{excluded} from model inputs to ensure code-grounded evaluation.
\end{itemize}

\paragraph{Why two tasks? From workflow to interface.}
Compliance reviews begin with identifying code locations that utilize APIs, permission gates, background services, storage, and transmission paths. This motivates a \emph{localization} view that evaluates whether a system can link dispersed signals to concrete anchors at three levels: file, module, and line (or short span). Once an anchor is identified, the task shifts to a content judgment over a specific code window: \emph{given this snippet, which provisions are implicated?} The \emph{judgment} view fixes the window and treats the outcome as a set of native identifiers. Separating these views (navigation vs.\ judgment) enhances interpretability, enables targeted ablations, and facilitates cross-jurisdiction comparisons without conflating steps.

\begin{figure*}[t]
  \centering
  \includegraphics[width=1\textwidth]{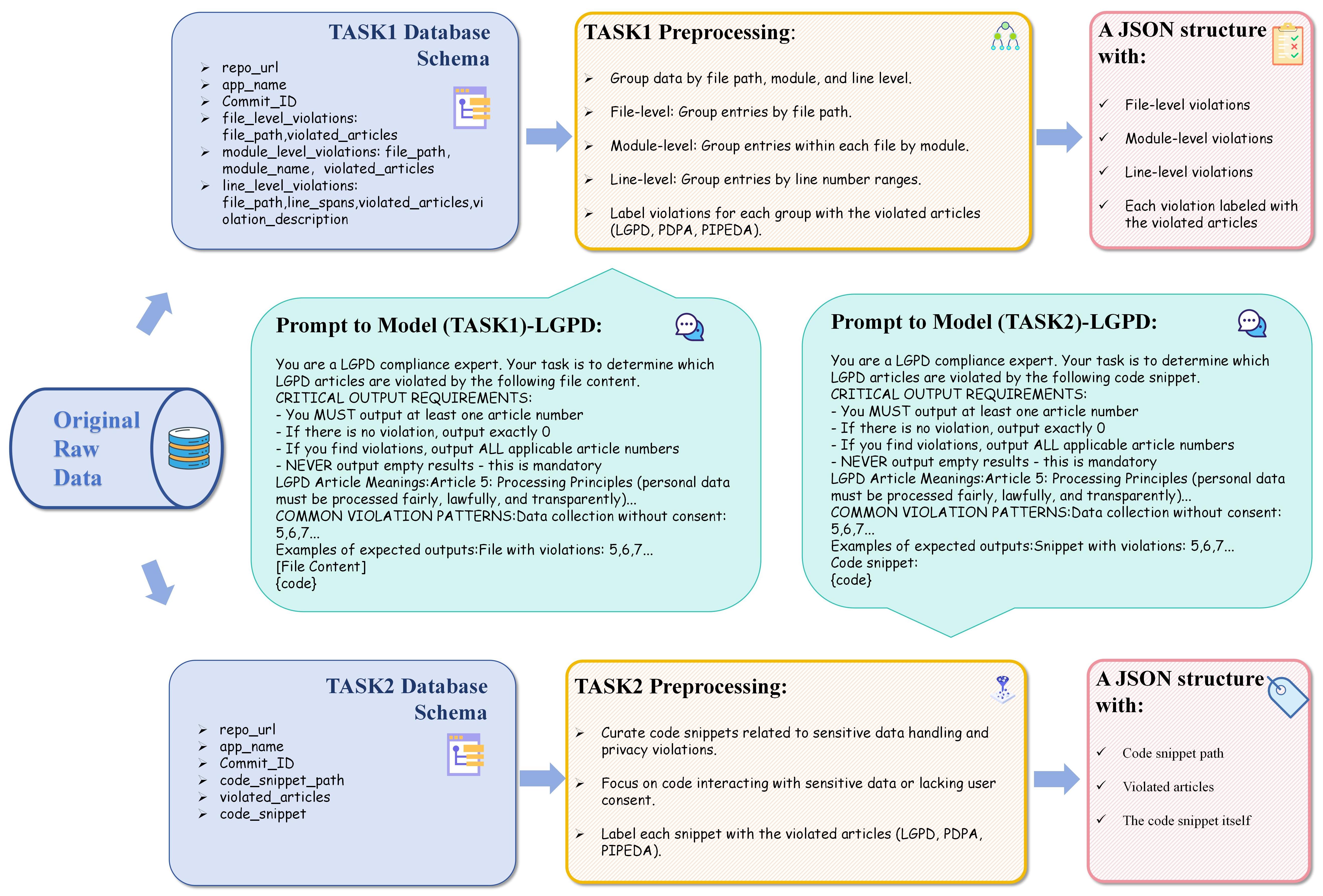}
  \caption{Task Design Overview. The figure illustrates the workflow for Task~1 (retrieval and localization) and Task~2 (multi-label judgment).}
  \label{fig:task-design}
\end{figure*}

To illustrate the overall workflow of both Task~1 (retrieval and localization) and Task~2 (multi-label judgment), see Figure~\ref{fig:task-design}. The figure provides a visual overview of how the tasks are structured, detailing the steps from data preprocessing to task execution.

\paragraph{Gold-corpus reshaping after task definition.}
After fixing task semantics, we deterministically reshape the raw artifact into two machine-readable views. No resampling, relabeling, or partitioning occurs; provenance (repository URL, app name, commit identifier) is retained verbatim. This prevents hyperparameter overfitting to small validation sets and aligns CompliBench with audit scenarios where models are used \emph{as-is}.

\paragraph{Granularity semantics for localization.}
We define three anchor levels for Task~1 to match how violations surface in code:
\begin{itemize}[leftmargin=1.35em, itemsep=0.22em, topsep=0.05em]
  \item \textbf{File.} A source file concentrating on privacy-relevant behaviors (e.g., a manager aggregating identifiers or orchestrating uploads). File anchors prioritize breadth over pinpoint precision.
  \item \textbf{Module.} A file-scoped logical unit, typically a top-level class or component that binds responsibilities (e.g., a service or activity). Module anchors assess whether a system can narrow signals to the responsible unit.
  \item \textbf{Line or short span.} A concrete statement or compact range that exhibits the behavior (e.g., persistent storage writes, unguarded broadcasts, or network transmission of sensitive fields). These anchors prioritize precision and provide strong evidence for subsequent judgment.
\end{itemize}
Anchors are mutually reinforcing: models generalizing well at coarse levels but failing at finer levels are of limited utility; conversely, line-level precision without broader context often leads to poor recall.

\paragraph{Snippet window policy for judgment.}
Task~2 standardizes the code window used for legal judgment. These windows include the focal lines and a small, deterministic context for self-containment (e.g., imports and surrounding signatures when necessary). We deduplicate only when content matches byte-for-byte \emph{and} the pointer remains consistent; otherwise, distinct anchors are preserved to maintain audit trails.

\paragraph{Identity, normalization, and de-duplication.}
To ensure stability across regenerations:
\begin{enumerate}[leftmargin=1.6em, itemsep=0.22em, topsep=0.05em]
  \item \emph{Identity.} Task~1 identities are tuples (file, app, repository, commit). Task~2 identities are (file, line or span, commit).
  \item \emph{Normalization.} Paths use project-relative POSIX separators; module names default to the primary top-level class/component, derived deterministically from the path; line anchors are recorded as closed spans when needed.
  \item \emph{De-duplication.} Task~1 unions repeated evidence into section-level sets; Task~2 collapses exact duplicates with consistent pointers but retains content-identical snippets mapped to different pointers (to preserve auditability across call sites).
\end{enumerate}

\paragraph{Shaping rules from the raw artifact.}
Each validated record is represented as $(r,a,c,p{:}\ell,s,A)$, where $r$ is the repository, $a$ is the app, $c$ is the commit, $p{:}\ell$ represents the pointer (file and line or span), $s$ is the snippet, and $A$ is the set of native article identifiers.
\begin{itemize}[leftmargin=1.35em, itemsep=0.22em, topsep=0.05em]
  \item \textbf{Task~1 (retrieval \& localization).} Group by (file, app, repository, commit). Within each file record, produce three sections—file, module, line—and aggregate implicated native articles per section. Module keys are derived deterministically (e.g., top-level class/component) to avoid heuristic variance.
  \item \textbf{Task~2 (multi-label judgment).} Group by the stable snippet pointer (file and line/span at commit). Keep one code window per pointer and union all implicated native articles into a set.
\end{itemize}
Both views preserve jurisdiction-native identifiers (LGPD, PDPA, PIPEDA), enabling cross-law analysis while maintaining doctrinal traceability.

\paragraph{Resulting split of the raw dataset (non-training).}
The reshaping yields two distinct \emph{views} over the same gold corpus rather than train/dev/test partitions. Each validated instance contributes (i) to the Task~1 file record that owns its path, enriching one or more of the three sections, and/or (ii) to the unique Task~2 snippet keyed by its pointer. No instance is held out for training, and no labels are synthesized during shaping.

\medskip
\noindent\textbf{Task artifacts: compact schemas.} We release one artifact per task and jurisdiction.

\paragraph{Edge cases and policy decisions.}
We standardize several edge cases to maintain stability in task views: (i) exclude files with unstable anchors (generated/obfuscated) unless pointers can be re-validated; (ii) for cross-file flows, retain per-file anchors to avoid ambiguous ownership; (iii) when module names are ambiguous (e.g., companion objects), default to the top-level unit that best matches the file name; (iv) merge overlapping spans only if their article sets are identical; otherwise, preserve them as distinct.

\paragraph{Benefits of the chosen design.}
The two-view formulation offers (i) \emph{operational realism} by aligning with audit workflow; (ii) \emph{diagnostic clarity} by separating navigation from judgment; (iii) \emph{granularity sensitivity} by enforcing consistency from file to line; (iv) \emph{jurisdictional traceability} by preserving native identifiers; and (v) \emph{reproducibility} through deterministic normalization, pointer validation, and lossless reshaping. These properties enable principled comparisons across laws and repositories while maintaining fidelity to real-world review practices.

\medskip
\noindent
Having finalized the task semantics and shaping rules, the next step is to determine \emph{how performance should be measured}. Without carefully designed metrics, the two views would remain descriptive rather than evaluative. Section~\ref{sec:metrics} introduces standard measures alongside stability-aware composites that capture brittleness, cross-granularity inconsistency, and cross-jurisdiction variability—key factors in real audits as much as raw predictive power.

\subsection{Evaluation Metrics}
\label{sec:metrics}

\noindent
The two tasks in \emph{CompliBench} require distinct measurement lenses. We therefore report (i) \emph{task–specific base metrics} drawn from information retrieval and multi-label learning, and (ii) \emph{stability–aware composites} that summarize performance across granularities, metrics, and jurisdictions while penalizing brittleness. Our base choices follow standard IR and classification practice \cite{manning2008ir,jarvelin2002ndcg,zhang2014multilabel}, and the composite constructions build on classical multi-criteria ideas (TOPSIS, Mahalanobis) \cite{hwang1981topsis,mahalanobis1936}. All definitions below match the public implementation.

\paragraph{Task~1: retrieval and localization (file, module, line).}
For each key we evaluate the ranked list of article identifiers against the gold set. Matching can be strict (full pointer) with an optional relaxed diagnostic that falls back to the file path; metrics are computed per key and then averaged. Higher is better for all base metrics in Table~\ref{tab:task1-base}. Rank-sensitive measures (MRR, MAP, nDCG) follow established IR definitions \cite{manning2008ir,jarvelin2002ndcg}.

\begin{table*}[t]
  \centering
  \caption{Task~1 base metrics and closed–form definitions. $G$ = gold set, $P$ = ranked predictions, $|G|=R$, $\mathbb{1}[\cdot]$ indicator, $\operatorname{pos}(x)$ the rank position of $x$.}
  \label{tab:task1-base}
  \begin{tabularx}{\linewidth}{@{}l X p{0.35\linewidth}@{}}
    \toprule
    \textbf{Metric} & \textbf{Formula} & \textbf{Interpretation} \\
    \midrule
    Acc@1 & $\displaystyle \frac{|G \cap P_{[:1]}|}{|G|}$ & Recall at rank~1. \\
    Acc@5 & $\displaystyle \frac{|G \cap P_{[:5]}|}{|G|}$ & Recall at rank~5. \\
    R–precision & $\displaystyle \frac{|G \cap P_{[:R]}|}{R}$ & Precision at $R{=}|G|$. \\
    MRR & $\displaystyle \frac{1}{\min\{\operatorname{pos}(x): x\in P\cap G\}}$ & Reciprocal rank of the first correct prediction (zero if no hit). \\
    MAP & $\displaystyle \frac{1}{|G|}\sum_{x\in G}\frac{|\{y\in P: \operatorname{pos}(y)\le \operatorname{pos}(x),\, y\in G\}|}{\operatorname{pos}(x)}$ & Mean average precision aggregated over all relevant items. \\
    nDCG@5 & $\displaystyle \frac{\sum_{i=1}^{5}\frac{\mathbb{1}[P_i\in G]}{\log_2(i{+}1)}}{\sum_{i=1}^{\min(|G|,5)}\frac{1}{\log_2(i{+}1)}}$ & Normalized discounted cumulative gain at rank~5, comparing to the ideal ordering. \\
    \bottomrule
  \end{tabularx}
\end{table*}

\paragraph{Task~2: multi-label judgment at the snippet level.}
Each snippet corresponds to a set of violated articles. We report Micro/Macro/Weighted-F1, Jaccard similarity, Hamming loss, and normalized coverage error (Table~\ref{tab:task2-base}). These are standard in multi-label evaluation \cite{zhang2014multilabel,tsoumakas2010mlsurvey}; F1 traces to classic effectiveness measures \cite{vanRijsbergen1979}, Jaccard to set overlap, and Hamming loss to bit-error accounting \cite{hamming1950}. For F1 and Jaccard, larger is better; for Hamming loss and normalized coverage error, smaller is better. The implementation supports treating empty–empty pairs as a perfect Jaccard match and uses a stable, prediction-first ordering to build the label ranking for coverage error.

\begin{table*}[t]
  \centering
  \caption{Task~2 base metrics. $Y$ and $\hat{Y}$ are binary label matrices; $y_i$ and $\hat{y}_i$ are the true and predicted sets for sample $i$; $L$ is the number of labels; $\vert\cdot\vert$ denotes set size.}
  \label{tab:task2-base}
  \begin{tabularx}{\linewidth}{@{}l X X@{}}
    \toprule
    \textbf{Metric} & \textbf{Formula} & \textbf{Interpretation} \\
    \midrule
    Micro–F1 & $\displaystyle \frac{2\,\mathrm{TP}}{2\,\mathrm{TP}+\mathrm{FP}+\mathrm{FN}}$ & Global F1 over all labels and samples. \\
    Macro–F1 & $\displaystyle \frac{1}{L}\sum_{\ell=1}^{L}\mathrm{F1}_\ell$ & Unweighted mean of per–label F1. \\
    Weighted–F1 & $\displaystyle \sum_{\ell=1}^{L}w_\ell\,\mathrm{F1}_\ell,\;w_\ell\propto\text{label frequency}$ & Frequency–weighted mean of per–label F1. \\
    Jaccard (samples) & $\displaystyle \frac{1}{N}\sum_{i=1}^N \frac{|y_i\cap \hat{y}_i|}{|y_i\cup \hat{y}_i|}$ & Per–sample set overlap; optional empty–empty=1. \\
    Hamming loss & $\displaystyle \frac{1}{NL}\sum_{i=1}^{N}\sum_{\ell=1}^{L}\mathbb{1}[Y_{i\ell}\neq \hat{Y}_{i\ell}]$ & Fraction of misassigned label bits (lower is better). \\
    Normalized coverage error & $\displaystyle \frac{1}{N}\sum_{i=1}^{N}\frac{\max_{t\in y_i}\operatorname{rank}_i(t)-1}{L-1}$ & How deep one must go in the label ranking to cover all true labels (lower is better). \\
    \bottomrule
  \end{tabularx}
\end{table*}

\paragraph{Why new composites?}
Traditional base metrics (e.g., MAP or F1) assess performance along a single slice and cannot reveal \emph{cross-granularity stability}, \emph{metric consistency}, or \emph{cross-jurisdiction robustness}. These dimensions matter operationally: auditors need models that retain quality from file to line, remain coherent across recall/precision/ranking axes, and behave predictably under different legal regimes. Our composites—SGS, RCS, CRGS, and OCS—make these desiderata explicit.

\subsubsection*{SGS: cross-granularity stability for Task~1}
\emph{Question.} Is retrieval quality \emph{consistently} good from file to line, or does it spike at one level and collapse at another?  
\emph{Rationale.} Auditors must triage broadly and then pinpoint precisely; the weakest level often governs utility.  
\emph{Definition.} For Task~1 metric $m$ (e.g., Acc@1), let $v_{m,\ell}\in[0,1]$ be the score at level $\ell\in\{\text{file, module, line}\}$. Define
\begin{equation}
\label{eq:sgs}
\mathrm{SGS}(m)=
\bigg(\frac{1}{3}\sum_{\ell}(v_{m,\ell}{+}\varepsilon)^{-1}\bigg)^{-1}
\cdot
\exp\!\Big(-\alpha\,\mathrm{CV}^2(\{v_{m,\ell}\})\Big),
\end{equation}
where $\varepsilon$ prevents division by zero, $\mathrm{CV}$ is the coefficient of variation, and $\alpha{=}1$ by default. The first factor is the harmonic mean (punishing any weak level); the exponential penalizes cross-level volatility.  
\emph{Orientation.} Larger is better: higher SGS indicates the model maintains consistent ranking quality across granularities, rather than overfitting to only one anchor level.

\subsubsection*{RCS: regulation-wise composite per task}
\emph{Question.} Given several base signals for one task under one law, how close is a model to the \emph{ideal} system?  
\emph{Rationale.} Different metrics correlate; naive averaging double-counts redundancy. We therefore adopt a TOPSIS-style distance-to-ideal that is covariance-aware \cite{hwang1981topsis,mahalanobis1936}.  
\emph{Definition.} Represent a model by $\mathbf{z}\in[0,1]^K$ where all $K$ metrics are “higher is better”. For Task~1 we use the six $\mathrm{SGS}(m)$ values; for Task~2 we use
\[
  [\text{micro-F1},\text{macro-F1},\text{weighted-F1},\text{Jaccard},\,1{-}\text{NCE},\,1{-}\text{Hamming}].
\]
Let $C$ be the (ridge-regularized) covariance of metric columns and define
\begin{equation}
\label{eq:maha}
d_M(\mathbf{x},\mathbf{y})=\sqrt{(\mathbf{x}-\mathbf{y})^{\!\top}\,C^{-1}\,(\mathbf{x}-\mathbf{y})}.
\end{equation}
With $\mathbf{z}^{+}=\mathbf{1}$ and $\mathbf{z}^{-}=\mathbf{0}$, the TOPSIS score is
\begin{equation}
\label{eq:topsis}
\mathrm{RCS}=\frac{d_M(\mathbf{z},\mathbf{z}^{-})}{d_M(\mathbf{z},\mathbf{z}^{+})+d_M(\mathbf{z},\mathbf{z}^{-})}\in[0,1].
\end{equation}
\emph{Orientation.} Larger is better. Lower values indicate contradictory signals across metrics (e.g., strong precision but very low recall); high RCS evidences internal consistency within a task and regulation.

\subsubsection*{CRGS: cross-regulation geometric score}
\emph{Question.} Does the model remain competitive across jurisdictions, rather than excelling on one and failing on others?  
\emph{Rationale.} Real deployments rarely target a single law; volatility across regimes undermines trust.  
\emph{Definition.} Let $\{r_\rho\}$ be RCS values over $\rho\in\{\text{LGPD, PDPA, PIPEDA}\}$. Define
\begin{equation}
\label{eq:crgs}
\mathrm{CRGS}
=
\bigg(\prod_{\rho}\max(r_\rho,\varepsilon)\bigg)^{1/|\mathcal{R}|}
\cdot
\exp\!\Big(-\beta\,\mathrm{Var}(\{r_\rho\})\Big),
\end{equation}
with $\beta{=}2$.  
\emph{Orientation.} Larger is better. A single weak jurisdiction sharply lowers the geometric mean; instability further reduces the score, making CRGS a direct indicator of cross-law robustness.

\subsubsection*{OCS: overall composite with cross-task coupling}
\emph{Question.} Is a model jointly good at \emph{localization} and \emph{judgment}, and is that balance stable across laws?  
\emph{Rationale.} End-to-end audits require both capabilities; over-optimizing one at the expense of the other diminishes usefulness.  
\emph{Definition.} For each regulation $\rho$ let $(x_\rho,y_\rho)$ be T1–RCS and T2–RCS. Couple tasks within a law via
\begin{equation}
\label{eq:ocs-couple}
S_\rho=\frac{2\,x_\rho\,y_\rho}{x_\rho+y_\rho+\varepsilon}\cdot \exp\!\big(-\gamma\,|x_\rho-y_\rho|\big),
\end{equation}
then aggregate across laws as in CRGS:
\begin{equation}
\label{eq:ocs}
\mathrm{OCS}
=
\bigg(\prod_{\rho}\max(S_\rho,\varepsilon)\bigg)^{1/|\mathcal{R}|}
\cdot
\exp\!\Big(-\delta\,\mathrm{Var}(\{S_\rho\})\Big),
\end{equation}
with $\gamma{=}\delta{=}2$.  
\emph{Orientation.} Larger is better. High OCS means localization and judgment reinforce one another within each law and remain stable across laws.

\paragraph{Design choices and implications.}
\emph{Harmonic means} bound scores by the weakest component, reflecting pipeline fragility at the weakest granularity (SGS) or task (OCS).  
\emph{Covariance-aware ranking} (Mahalanobis–TOPSIS) avoids double-counting correlated metrics and yields a principled distance-to-ideal \cite{hwang1981topsis,mahalanobis1936}.  
\emph{Geometric fusion across laws} emphasizes uniform competence: a single failure should drag down the composite.  
\emph{Explicit stability penalties} make volatility visible and appropriately discounted. Defaults ($\alpha{=}1,\beta{=}2,\gamma{=}2,\delta{=}2$, ridge $0.1$) match the released evaluator.

\paragraph{Reporting protocol.}
For Task~1 we publish per-level base metrics, per-metric SGS, and T1–RCS by regulation; for Task~2 we publish base metrics and T2–RCS; we then report CRGS per task and a single OCS number. The tool additionally prints strict/relaxed key-matching diagnostics for Task~1 to make ceiling effects explicit.

\section{Experimental Setup}
\label{sec:setup}

\noindent
This section specifies the conditions under which we evaluate models on \emph{CompliBench}. We describe the datasets and snapshot policy, the model cohort, the inference protocol (prompts and hyperparameters), and the scoring configuration. Our design aims for \emph{reproducibility} and \emph{auditability} under a realistic setting where models are accessed uniformly through API endpoints rather than locally fine-tuned instances, a choice aligned with current best practices for controlled empirical reporting in software/ML systems \cite{Pineau2021Reproducibility,Henderson2018Robustness}.

\paragraph{Corpora and snapshots.}
We evaluate on the expert-validated gold corpus introduced in Sections~\ref{sec:dataset} and~\ref{sec:task-design}. The corpus comprises Android projects with Kotlin/Java/XML sources and jurisdiction-native labels (LGPD, PDPA, PIPEDA). Each instance stores its repository URL, application identifier, and an immutable commit identifier; all pointers (file and line/span) are resolved at that commit to eliminate drift. The benchmark uses the \emph{entire} gold corpus for evaluation—no train/dev/test splits—and all task-specific artifacts are deterministically shaped from this raw set (cf.\ Tables~\ref{tab:task1-schema}–\ref{tab:task2-schema}). We report per-jurisdiction results and aggregate using the stability-aware composites in Section~\ref{sec:metrics}. This snapshot discipline mirrors artifact-evaluation norms for repeatability and exact provenance recovery \cite{Pineau2021Reproducibility}.

\paragraph{Model cohort.}
To probe generality across modeling families while keeping access uniform, we evaluate six widely used LLMs via API endpoints under a single invocation protocol. Table~\ref{tab:models} reports, for each model, whether its weights are openly available for self-hosting and a short descriptive note. All systems are treated as black-box text-in/text-out services under a common protocol, ensuring comparability across proprietary and open-weight families. No model is fine-tuned or adapted on \emph{CompliBench}; the benchmark assesses \emph{intrinsic} capabilities only.

\begin{table}[H]
  \centering
  \caption{Models evaluated in CompliBench. All models were accessed via API; openness indicates whether public weights are available for self-hosting.}
  \label{tab:models}
  \begin{tabularx}{\linewidth}{@{}l c X@{}}
    \toprule
    \textbf{Model} & \textbf{Open?} & \textbf{Notes} \\
    \midrule
    gpt-4o & No & General-purpose; strong code/\& compliance reasoning; robust instruction-following. \\
    o1 & No & Reasoning-oriented family variant; optimized for stepwise inference and tool-use style outputs. \\
    claude-3.5-sonnet-20241022 & No & Sonnet series (Oct 2024); long-context, safety-tuned, strong summarization/extraction. \\
    claude-3.7-sonnet-20250219 & No & Sonnet series (Feb 2025); improved reasoning and red-teaming resilience. \\
    gemini-2.5-pro & No & Multimodal-capable; competitive code understanding and retrieval-style outputs. \\
    qwen2.5-72b-instruct & Yes & 72B open-weight instruction model; strong code grounding and list-style responses (evaluated via API). \\
    \bottomrule
  \end{tabularx}
\end{table}

\noindent
For comparability, all models receive the same inputs and decoding settings. This uniform, API-based evaluation avoids deployment-pipeline confounds and emphasizes stability and fairness across families.

\paragraph{Prompting and jurisdiction-specific instructions.}
Inference follows a fixed instruction template per jurisdiction. Each template (i) summarizes the scope of the law, (ii) enumerates decision cues (consent, notice, collection, retention, security, transfer), (iii) lists legitimate exceptions, and (iv) \emph{constrains outputs to native article identifiers only}. For Task~1, the prompt emphasizes linking the given file/module/line anchor to implicated provisions; for Task~2, the prompt fixes the snippet window and requests the full set of implicated provisions. Any free-text rationale produced by a model is ignored during scoring; only identifiers are parsed. We standardize and constrain prompts to minimize confounding effects of uncontrolled free-text generation and to enforce comparable, structured outputs across models and jurisdictions, consistent with guidance in recent prompting surveys on reducing prompt-induced variance \cite{Liu2023PromptSurvey,Dong2023ICLSurvey}.

\paragraph{Decoding, retries, and parallelism.}
All API calls use identical decoding and retry settings:
\begin{itemize}[leftmargin=1.4em, itemsep=0.25em, topsep=0.15em]
  \item \emph{Temperature} $=0.0$ (deterministic decoding; eliminates sampling stochasticity so that observed differences reflect model competence rather than random draws), \emph{max tokens} $=2048$.
  \item \emph{Timeout} $=180$ seconds per request; up to $3$ retry attempts on failure.
  \item \emph{Concurrency} via a thread-pool executor sized to the number of models, executing runs in parallel over the same instances; this provides fair wall-clock comparison while mitigating provider-specific throttling artifacts.
\end{itemize}
These controls attribute outcome differences to models (not decoding noise), bound latency, and remain robust to transient API failures. All requests are logged with timestamps and model identifiers; a monitoring thread records per-model progress, supporting post-hoc reproducibility checks \cite{Pineau2021Reproducibility,Henderson2018Robustness}.

\paragraph{Task-specific inputs.}
For \emph{Task~1} (retrieval and localization), each instance provides the repository URL, application identifier, commit identifier, the target granularity (file/module/line), the key (file path, module name, or line/span), and a compact context (neighboring lines or structural cues). Models return a ranked or unordered set of native article identifiers per key; the evaluator derives Acc@k, MRR, MAP, and nDCG@5 from rankings and also accepts unordered sets.  
For \emph{Task~2} (multi-label judgment), each instance provides a stable pointer (file and line/span) and the code window; models return a set of native identifiers for that snippet. In both tasks, expert notes are \emph{never} shown to models.

\paragraph{Scoring configuration.}
Unless otherwise stated, scoring uses \emph{strict} key matching (full pointer equality) for Task~1. A \emph{relaxed} diagnostic that falls back to file-level matching can be enabled to quantify coverage ceilings on noisy keys; it does not affect primary reported scores. Task~2 adopts conventional multi-label measures (micro/macro/weighted F1, Jaccard over samples, Hamming loss, normalized coverage error) as consolidated in Section~\ref{sec:metrics}. Composite scores follow Section~\ref{sec:metrics}: SGS (harmonic mean with cross-granularity variance penalty), RCS (TOPSIS with Mahalanobis geometry), CRGS (geometric mean with cross-law variance penalty), and OCS (harmonic coupling of Task~1 and Task~2 per law, then CRGS across laws). Hyperparameters use the reference defaults.

\paragraph{Prediction parsing and validation.}
Outputs are parsed to extract native identifiers only; any extraneous text is discarded. For Task~1, duplicate matches within a key are deduplicated before computing ranking metrics; for Task~2, sets are binarized against the jurisdiction’s label universe. The evaluator emits per-task diagnostics—matched keys, unmatched anchors (strict vs.\ relaxed), label cardinalities, and per-law confusion summaries—to support ablations and error analysis.

\paragraph{Reproducibility.}
We release prompts, scripts, and configuration. Determinism is enforced by: (i) immutable commit identifiers, (ii) temperature $0$ decoding, (iii) a uniform prompting protocol across models and tasks, and (iv) fixed composite-aggregation parameters. The harness writes machine-readable result files (per-task base metrics, per-metric SGS, per-task RCS, per-task CRGS, overall OCS) and a human-readable log summarizing strict/relaxed coverage, following emerging transparency norms for computational experiments \cite{Pineau2021Reproducibility}.

\paragraph{Hyperparameters and execution controls.}
For transparency, Table~\ref{tab:hparams} lists the inference and control settings applied to all runs.

\begin{table}[H]
  \centering
  \caption{Uniform inference hyperparameters and execution controls.}
  \label{tab:hparams}
  \begin{tabularx}{\linewidth}{@{}l l X@{}}
    \toprule
    \textbf{Setting} & \textbf{Value} & \textbf{Purpose} \\
    \midrule
    Temperature & 0.0 & Deterministic decoding across models. \\
    Max tokens & 2048 & Room for article lists and minimal rationale. \\
    Timeout & 180\,s & Bound per-call latency on slower providers. \\
    Retries & 3 & Tolerate transient API/network failures. \\
    Concurrency & \#models & Parallel per-model execution via thread pool. \\
    Prompting & Fixed per law & Enforce native IDs, list-only outputs. \\
    Key matching (Task~1) & Strict (default) & Relaxed file-level fallback for diagnostics only. \\
    Composite params & $\alpha{=}1,\ \beta{=}2,\ \gamma{=}2,\ \delta{=}2$ & Stability-aware aggregation defaults. \\
    TOPSIS ridge & 0.1 & Covariance regularization for Mahalanobis geometry. \\
    \bottomrule
  \end{tabularx}
\end{table}
\section{Results and Analysis}
\label{sec:results}

\noindent
This section reports empirical findings on \emph{CompliBench}. We proceed in two layers. First, we examine \emph{task–level} performance with radar plots that place multiple base metrics onto a single canvas, allowing us to reason about \emph{shape}, \emph{balance}, and \emph{consistency} at a glance without flooding the page with numbers. Second, we will consolidate these signals using the stability–aware composites introduced in \S\ref{sec:metrics}, which quantify robustness across granularities and jurisdictions.

\subsection{Task–Level Results}
\label{sec:results-task}

\noindent
We begin with visual summaries of the base metrics. For Task~1 (retrieval and localization), each model has three radar panels—file, module, and line—whose axes are Acc@1, Acc@5, R–Precision, MRR, MAP, and nDCG@5 (definitions in \S\ref{sec:metrics}). The three colored polylines represent LGPD, PDPA, and PIPEDA. For Task~2 (snippet–level multi–label judgment), each model has one radar panel with Micro–F1, Macro–F1, Weighted–F1, Jaccard, $1-$Hamming, and $1-$NCE (we plot $1-$NCE so that “larger is better” is visually uniform). In all panels, \emph{larger and more regular} polygons connote stronger and more balanced performance; thin or skewed shapes reveal trade–offs or brittleness.

\begin{figure*}[t]
  \centering
  \includegraphics[width=\textwidth]{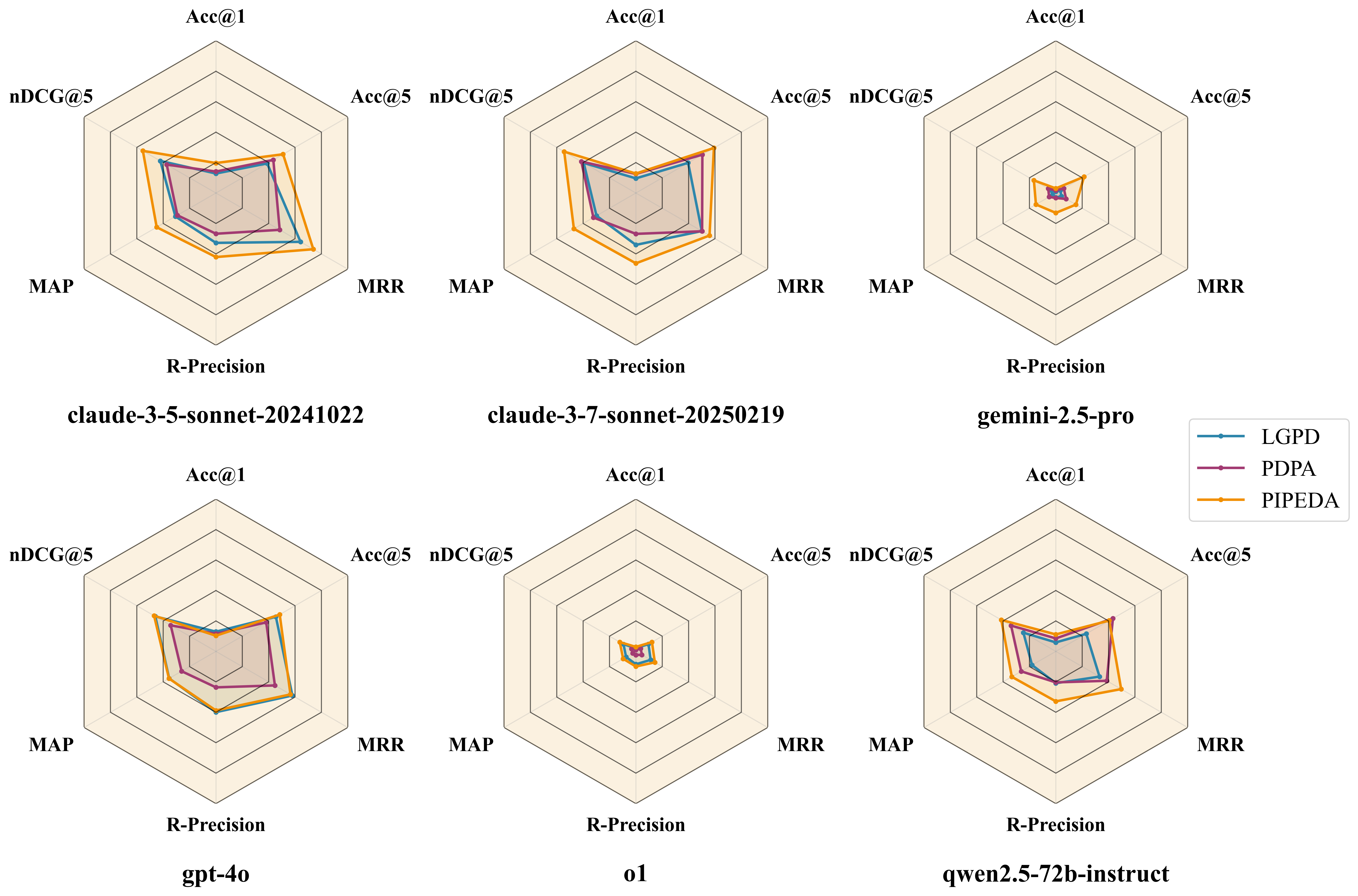}
  \caption{\textbf{Task~1 (file level).} Radar plots per model. Axes: Acc@1, Acc@5, R–Precision, MRR, MAP, nDCG@5. Larger, more regular polygons indicate stronger and more balanced retrieval at the file level.}
  \label{fig:t1-file}
\end{figure*}

\paragraph{Task~1 — File level.}
Figure~\ref{fig:t1-file} shows the file–level view. Three phenomena recur.

\emph{(i) Breadth versus early precision.} Many panels show “back–heavy” polygons: Acc@5 and nDCG@5 spokes extend further than Acc@1 and MRR. Systems frequently \emph{include} correct provisions in the top–$k$ set yet \emph{fail} to rank them first. When all three jurisdictional polylines share this wedge, the issue is largely ranking; when only one jurisdiction bows outward on Acc@5 but not on MRR, the tension appears regime–specific (label–space or taxonomy effects).

\emph{(ii) R–Precision as a cardinality check.} R–Precision typically sits between Acc@1 and Acc@5, as expected at depth $R{=}|G|$. When polygons hug the center on the R–Precision spoke despite a sizable Acc@5, we observe \emph{cardinality mismatch}: over– or under–prediction relative to the gold set diminishes precision-like measures even when breadth looks adequate.

\emph{(iii) Jurisdictional spread and crossings.} In some models one colored polyline consistently encloses the others (stable relative difficulty across metrics); in others the polylines cross repeatedly, indicating a metric-dependent ordering of jurisdictions—e.g., better breadth (Acc@5) for one law but better early precision (MRR) for another.

\begin{figure*}[t]
  \centering
  \includegraphics[width=\textwidth]{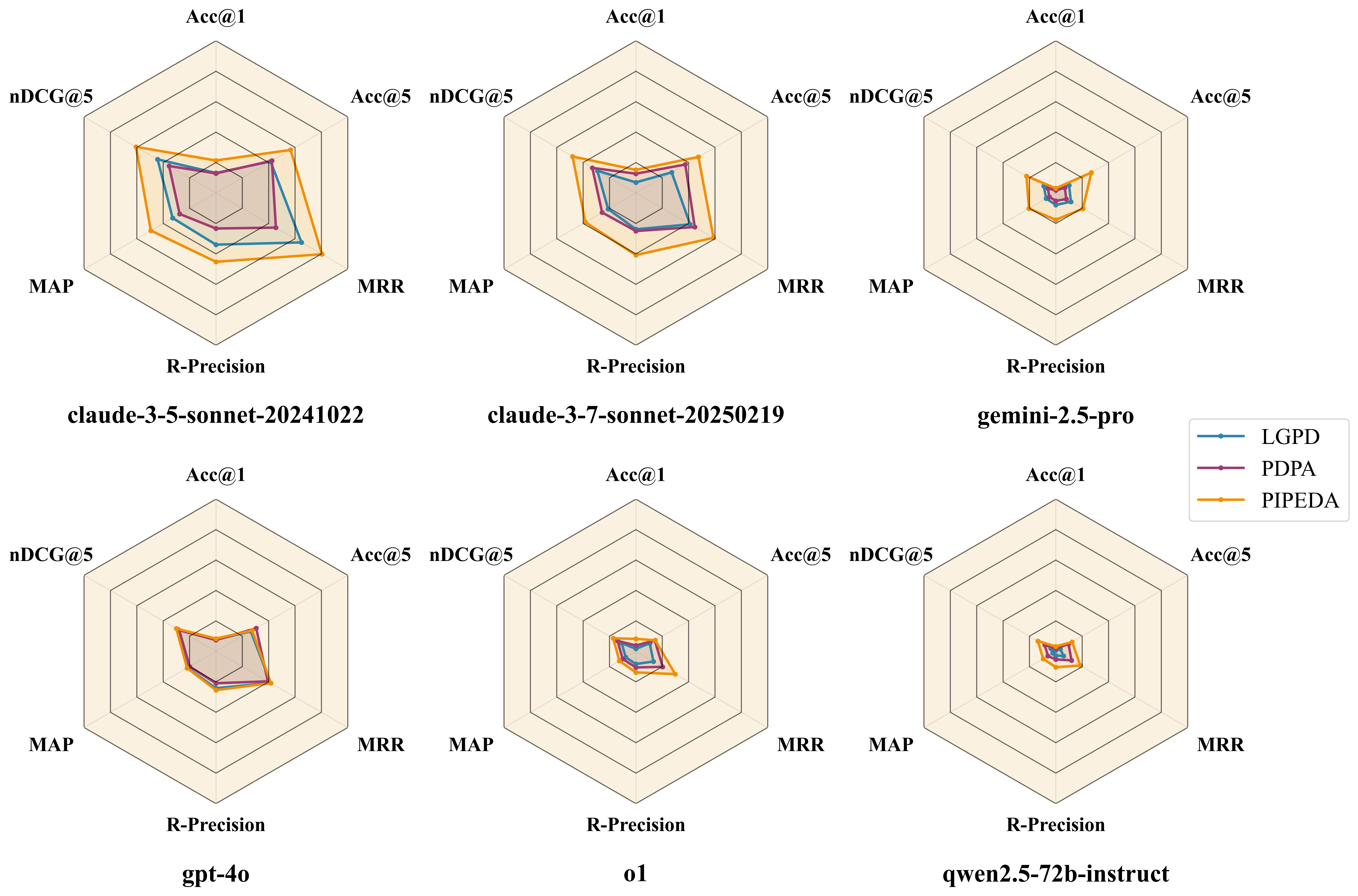}
  \caption{\textbf{Task~1 (module level).} Compared with file level, many models contract on MRR/Acc@1 while preserving Acc@5, revealing difficulties in precise ordering at component granularity.}
  \label{fig:t1-module}
\end{figure*}

\paragraph{Task~1 — Module level.}
Figure~\ref{fig:t1-module} stresses how zooming from a file to its principal component affects ranking stability.

\emph{(i) Contraction on early–precision spokes.} Relative to file level, MRR and Acc@1 spokes often shorten while Acc@5 holds steady. The polygon “flattens” toward early–precision axes: systems still surface the right provisions among a few candidates but become brittle in ordering them within the component boundary.

\emph{(ii) Persistence—or inversion—of jurisdiction ordering.} It is common to see the jurisdiction that led at file level \emph{not} leading at module level; colored polylines swap enclosure. Such inversions hint that certain regimes manifest violations more \emph{dispersed across files} (favoring file–level breadth) whereas others cluster within a recognizable unit (favoring module–level precision).

\emph{(iii) Shape regularity as robustness.} Where polygons remain roughly regular hexagons after the zoom, ability is retained uniformly; kite– or arrow–like shapes betray asymmetric degradation—either breadth without ranking or ranking without breadth.

\begin{figure*}[t]
  \centering
  \includegraphics[width=\textwidth]{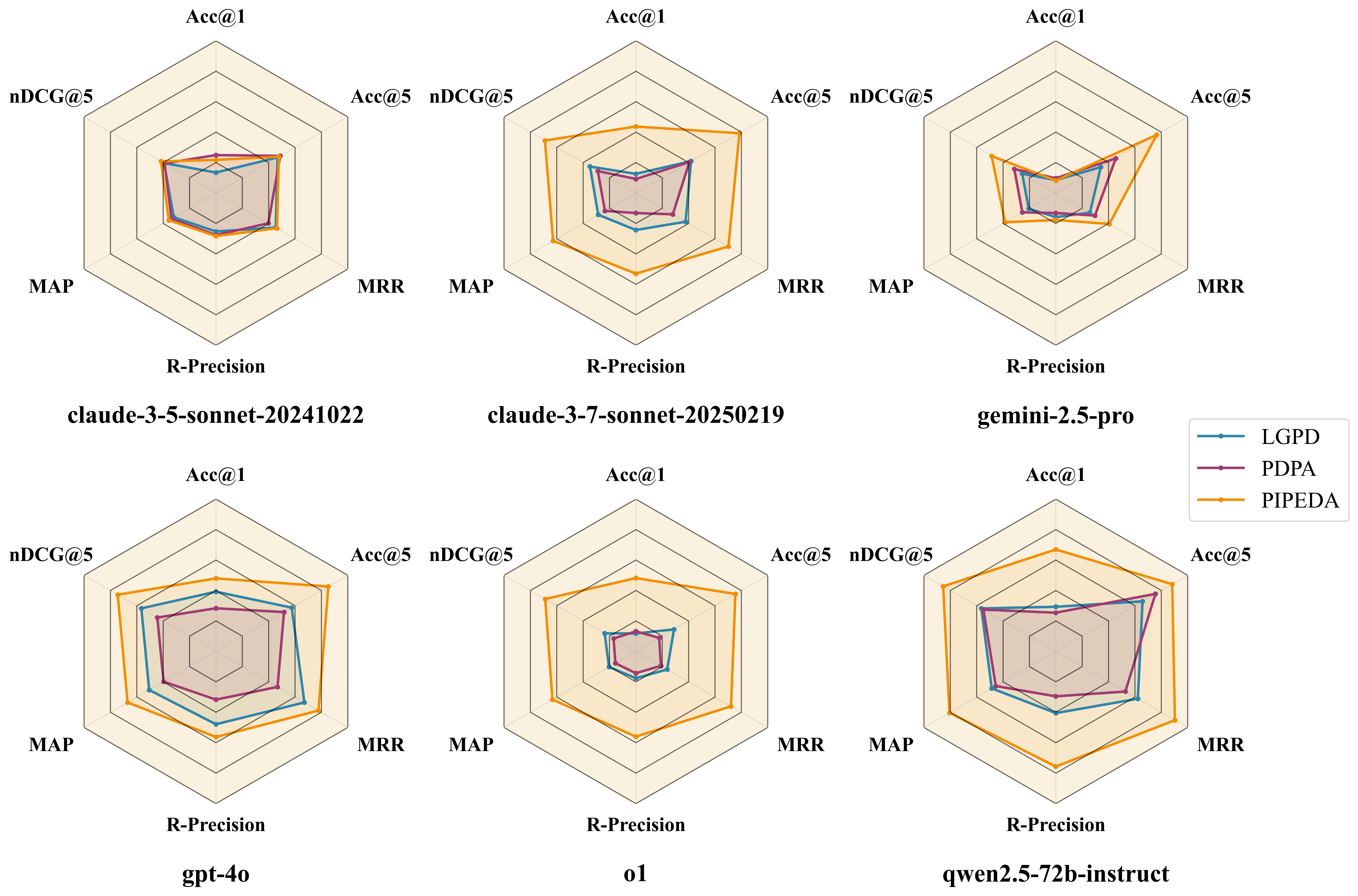}
  \caption{\textbf{Task~1 (line level).} Polygons often dilate relative to coarser levels, especially on ranking–sensitive axes (MRR, MAP, nDCG@5), reflecting the benefit of precise anchors.}
  \label{fig:t1-line}
\end{figure*}

\paragraph{Task~1 — Line level.}
Figure~\ref{fig:t1-line} presents the line–level view, where anchors are concrete statements or short spans.

\emph{(i) Granularity–driven dilation.} Polygons frequently \emph{dilate} on MRR, MAP, and nDCG@5 while maintaining or slightly improving Acc@5. Precise anchors reduce ambiguity; ranking–sensitive axes improve in tandem with breadth.

\emph{(ii) Residual irregularity and article disambiguation.} Some panels show a “tapered fan”: Acc@5 grows but MAP lags, meaning relevant provisions are \emph{found} yet not consistently \emph{ordered} ahead of distractors. If this holds across jurisdictions, the issue is generic ranking; if localized to one regime, provision granularity or co–occurrence is the likely culprit.

\emph{(iii) Cross–jurisdiction convergence at fine anchors.} The three polylines often draw closer at line level than at file/module levels. Once the behavior is pinned to a precise call site, the legal mapping depends less on repository layout and more on snippet semantics, narrowing jurisdictional gaps.

\paragraph{Cross–granularity diagnostics within a model.}
Placing the three panels of a given model side–by–side supports quick diagnosis:
\begin{itemize}[leftmargin=1.35em, itemsep=0.25em, topsep=0.1em]
  \item \textbf{Monotone expansion.} If polygons expand from file$\rightarrow$module$\rightarrow$line while preserving shape, the model scales uniformly with finer anchors—ideal for audits that need both triage and pinpointing.
  \item \textbf{Breadth–only gains.} Acc@5 rises but MRR/MAP stagnate: good for enumerating candidates, weaker for prioritization.
  \item \textbf{Ranking–only gains.} MRR/MAP rise but Acc@5 is flat: quick confirmation once a candidate set is known, but recall remains a concern.
  \item \textbf{Jurisdiction order flips.} Color order changes across granularities imply law–specific structural manifestations in code, precisely the instability our cross–law composite discounts (\S\ref{sec:metrics}).
\end{itemize}

\begin{figure*}[t]
  \centering
  \includegraphics[width=\textwidth]{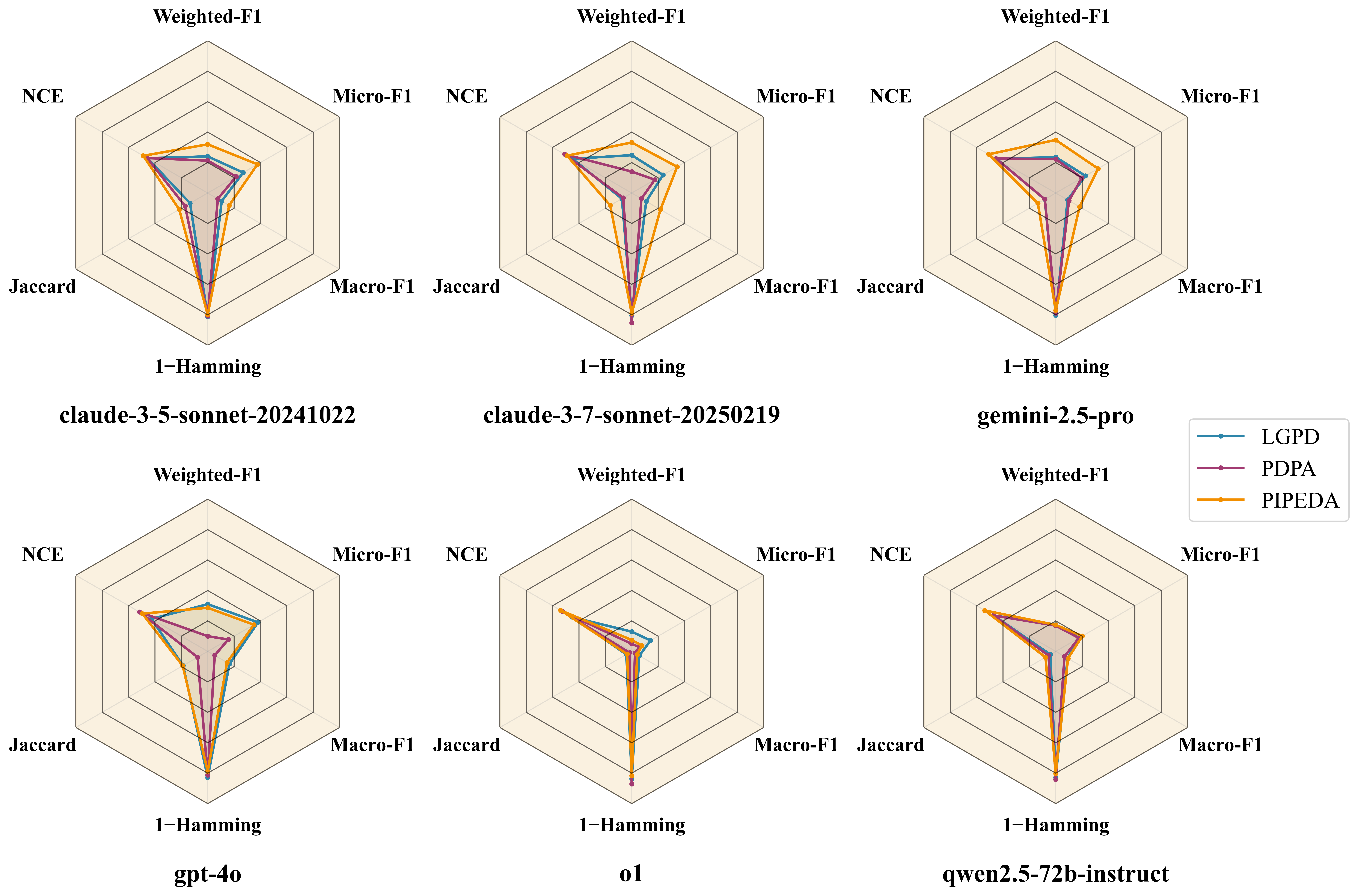}
  \caption{\textbf{Task~2 (snippet–level judgment).} Axes: Micro–F1, Macro–F1, Weighted–F1, Jaccard, $1-$Hamming, $1-$NCE. Triangle–like shapes leaning toward $1-$Hamming indicate majority–label dominance; rounded fans indicate better balance across common and rare provisions.}
  \label{fig:t2}
\end{figure*}

\paragraph{Task~2 — Snippet–level multi–label judgment.}
Figure~\ref{fig:t2} shows the judgment view. The axes capture complementary facets: Micro–F1 (global balance), Macro–F1 (rare labels), Weighted–F1 (frequency–weighted balance), Jaccard (set overlap), $1-$Hamming (bit–wise correctness), and $1-$NCE (depth needed to cover all truths).

\emph{(i) Triangle vs.\ fan.} Triangle–like shapes tilted toward the $1-$Hamming vertex with short Macro–F1 spokes indicate majority–label dominance and under–coverage of rare provisions. More rounded “fans” with long Jaccard and Macro–F1 spokes reflect better minority handling at a small cost to bit–wise accuracy.

\emph{(ii) Set–ranking agreement.} When Jaccard and $1-$NCE spokes extend together, the predicted sets and their induced rankings agree. Long Jaccard yet short $1-$NCE means the right sets are identified but some true labels sit deeper in the ranking—raising audit effort.

\emph{(iii) Jurisdictional alignment under fixed context.} Polylines for the three laws often overlap more in Task~2 than in Task~1 at coarse anchors. Fixed snippet context damps label–space variance; persistent gaps usually signal tightly coupled provisions unique to a regime.

\paragraph{Cross–law contrasts at fixed granularity.}
Within a panel (fixed model and granularity), differences among the three polylines diagnose regime–specific difficulty. If one polyline encloses the others across most axes, that law aligns more readily with the model’s priors. If polylines intersect—say, one leads on Acc@5 but lags on MRR—the law favors breadth but not early precision (or vice versa), informing whether a practitioner should rely on top–$k$ triage or demand higher first-rank correctness.

\paragraph{Cross–model contrasts at fixed law and granularity.}
Reading across panels row-wise, polygon \emph{area} orders overall strength while \emph{shape similarity} hints at behavioral similarity. Two similarly shaped polygons at different scales are workflow-substitutable (one is a stronger version of the other); very different shapes suggest complementary error profiles and motivate triage–then–judge pipelines.

\paragraph{Line-level (Task~1) versus snippet judgment (Task~2).}
Comparing a model’s line–level polygon with its snippet polygon reveals whether \emph{ranking gains from precise anchors} translate into \emph{consistent multi–label decisions}. Large, regular polygons in both views indicate that localization clarity is converted into sound judgment. If the line–level polygon is large but the snippet polygon remains triangle–like (high $1-$Hamming, low Macro–F1), the bottleneck lies in judgment, not localization.

\noindent
In summary: (i) many systems are breadth–positive yet ranking–fragile at coarse anchors; (ii) finer anchors generally dilate ranking–sensitive axes but do not universally fix article disambiguation; and (iii) jurisdictional gaps narrow with precise anchors yet persist where provisions are tightly coupled. These observations motivate the stability–aware composites in the next subsection, which explicitly discount brittle gains and reward uniformity across levels and laws (\S\ref{sec:metrics}).

\subsection{Cross–Jurisdiction and Cross–Task Comparisons}
\label{sec:results-xlaw-xtask}

\noindent
We now synthesize results \emph{across} jurisdictions (LGPD, PDPA, PIPEDA) and \emph{between} tasks (retrieval–localization vs.\ snippet judgment), using the stability-aware composites from Section~\ref{sec:metrics}. Specifically, we analyze per-law cross-granularity stability (\textbf{SGS}; Table~\ref{tab:sgs-by-law}), regulation-wise composites (\textbf{T1–RCS}, \textbf{T2–RCS}) along with their cross-law aggregates (\textbf{CRGS}; Tables~\ref{tab:t1-rcs}, \ref{tab:t2-rcs}), and the overall composite (\textbf{OCS}; Table~\ref{tab:ocs}).

\begin{table*}[t]
  \centering
  \caption{Task~1 cross–granularity stability (\textbf{SGS}) by law and model. For each law (row block) and each metric (column), the within–law maximum is \textbf{bold} and the minimum is \underline{underlined}.}
  \label{tab:sgs-by-law}
  \setlength{\tabcolsep}{4pt}
  \renewcommand{\arraystretch}{1.1}
  \resizebox{\textwidth}{!}{%
  \begin{tabular}{@{}llrrrrrr@{}}
    \toprule
    \textbf{Law} & \textbf{Model} & \textbf{SGS\_Acc@1} & \textbf{SGS\_Acc@5} & \textbf{SGS\_R-Prec} & \textbf{SGS\_MRR} & \textbf{SGS\_MAP} & \textbf{SGS\_nDCG@5} \\
    \midrule
    \multirow{6}{*}{LGPD}
      & claude-3.5-sonnet-20241022 & \textbf{0.1314} & \textbf{0.4185} & 0.2965 & \textbf{0.5520} & \textbf{0.3181} & \textbf{0.4175} \\
      & claude-3.7-sonnet-20250219 & 0.0870 & 0.3382 & 0.2590 & 0.4175 & 0.2531 & 0.3349 \\
      & gemini-2.5-pro              & 0.0104 & \underline{0.0358} & \underline{0.0386} & \underline{0.0547} & \underline{0.0293} & \underline{0.0383} \\
      & gpt-4o                      & 0.0798 & 0.3559 & \textbf{0.3209} & 0.5141 & 0.2865 & 0.3813 \\
      & o1                          & 0.0125 & 0.0952 & 0.0879 & 0.1294 & 0.0739 & 0.1021 \\
      & qwen2.5-72b-instruct        & \underline{0.0067} & 0.0411 & 0.0416 & 0.0878 & 0.0297 & 0.0518 \\
    \midrule
    \multirow{6}{*}{PDPA}
      & claude-3.5-sonnet-20241022 & \textbf{0.1441} & \textbf{0.4458} & \textbf{0.2562} & \textbf{0.4394} & \textbf{0.2972} & \textbf{0.3716} \\
      & claude-3.7-sonnet-20250219 & 0.1103 & 0.4146 & 0.1809 & 0.3648 & 0.2612 & 0.3291 \\
      & gemini-2.5-pro             & 0.0119 & \underline{0.0368} & \underline{0.0362} & \underline{0.0656} & \underline{0.0356} & \underline{0.0400} \\
      & gpt-4o                     & 0.0841 & 0.3653 & 0.2387 & 0.4316 & 0.2424 & 0.3300 \\
      & o1                         & \underline{0.0115} & 0.0586 & 0.0380 & 0.0734 & 0.0377 & 0.0531 \\
      & qwen2.5-72b-instruct       & 0.0221 & 0.1446 & 0.0819 & 0.1825 & 0.0905 & 0.1265 \\
    \midrule
    \multirow{6}{*}{PIPEDA}
      & claude-3.5-sonnet-20241022 & \textbf{0.2087} & 0.5125 & 0.3560 & 0.6009 & 0.4182 & 0.5005 \\
      & claude-3.7-sonnet-20250219 & 0.1275 & \textbf{0.5666} & \textbf{0.4558} & \textbf{0.6045} & \textbf{0.4545} & \textbf{0.5454} \\
      & gemini-2.5-pro             & 0.0302 & 0.2176 & 0.1551 & 0.1797 & 0.1798 & 0.1900 \\
      & gpt-4o                     & 0.0645 & 0.3578 & 0.3283 & 0.5159 & 0.2753 & 0.3867 \\
      & o1                         & \underline{0.0220} & \underline{0.0877} & \underline{0.0848} & \underline{0.1726} & \underline{0.0714} & \underline{0.1053} \\
      & qwen2.5-72b-instruct       & 0.0244 & 0.1652 & 0.1345 & 0.2587 & 0.1230 & 0.1841 \\
    \bottomrule
  \end{tabular}%
  }
\end{table*}

\begin{table}[t]
  \centering
  \caption{Task~1 regulation–wise composite (\textbf{T1–RCS}) and cross–law stability (\textbf{CRGS}). Maxima are \textbf{bold}; minima are \underline{underlined}.}
  \label{tab:t1-rcs}
  \begin{tabular}{@{}lrrrr@{}}
    \toprule
    \textbf{Model} & \textbf{LGPD} & \textbf{PDPA} & \textbf{PIPEDA} & \textbf{CRGS} \\
    \midrule
    claude-3.5-sonnet-20241022 & \textbf{0.3270} & \textbf{0.3100} & 0.3998 & \textbf{0.3425} \\
    claude-3.7-sonnet-20250219 & 0.2636 & 0.2673 & \textbf{0.4169} & 0.3054 \\
    gemini-2.5-pro             & \underline{0.0342} & \underline{0.0396} & 0.1576 & \underline{0.0594} \\
    gpt-4o                     & 0.3004 & 0.2737 & 0.3082 & 0.2936 \\
    o1                         & 0.0830 & 0.0468 & \underline{0.0976} & 0.0723 \\
    qwen2.5-72b-instruct       & 0.0466 & 0.1117 & 0.1532 & 0.0924 \\
    \bottomrule
  \end{tabular}
\end{table}

\begin{table}[t]
  \centering
  \caption{Task~2 regulation–wise composite (\textbf{T2–RCS}) and cross–law stability (\textbf{CRGS}). Maxima are \textbf{bold}; minima are \underline{underlined}.}
  \label{tab:t2-rcs}
  \begin{tabular}{@{}lrrrr@{}}
    \toprule
    \textbf{Model} & \textbf{LGPD} & \textbf{PDPA} & \textbf{PIPEDA} & \textbf{CRGS} \\
    \midrule
    claude-3.5-sonnet-20241022 & 0.3932 & \textbf{0.3811} & \textbf{0.4320} & \textbf{0.4011}\\
    claude-3.7-sonnet-20250219 & 0.3856 & 0.3624 & 0.4281 & 0.3905 \\
    gemini-2.5-pro             & 0.3821 & 0.3715 & 0.4133 & 0.3883 \\
    gpt-4o                     & \textbf{0.4336} & 0.3472 & 0.4171 & 0.3963 \\
    o1                         & \underline{0.3652} & \underline{0.3415} & \underline{0.3509} & \underline{0.3523} \\
    qwen2.5-72b-instruct       & 0.3697 & 0.3666 & 0.3718 & 0.3694 \\
    \bottomrule
  \end{tabular}
\end{table}

\begin{table}[t]
  \centering
  \caption{Overall composite (\textbf{OCS})—harmonic coupling of Task~1 and Task~2 per law with cross–law geometric aggregation. The maximum is \textbf{bold}; the minimum is \underline{underlined}.}
  \label{tab:ocs}
  \begin{tabular}{@{}lr@{}}
    \toprule
    \textbf{Model} & \textbf{OCS} \\
    \midrule
    claude-3.5-sonnet-20241022 & \textbf{0.3295} \\
    claude-3.7-sonnet-20250219 & 0.2919 \\
    gemini-2.5-pro             & \underline{0.0538} \\
    gpt-4o                     & 0.2736 \\
    o1                         & 0.0686 \\
    qwen2.5-72b-instruct       & 0.0853 \\
    \bottomrule
  \end{tabular}
\end{table}

\noindent
\textbf{Cross–jurisdiction patterns (Task~1).}
Several key observations emerge from Tables~\ref{tab:sgs-by-law} and~\ref{tab:t1-rcs}:
\begin{itemize}[left=0pt]
    \item \ding{182} \textbf{PIPEDA is the most stable regime for retrieval.} Across most models, PIPEDA yields the highest SGS values, particularly for models like \textbf{claude–3.5}. For instance, \textbf{claude–3.5} achieves SGS values of $0.2087$ (Acc@1), $0.5125$ (Acc@5), $0.6009$ (MRR), and $0.4182$ (MAP) on PIPEDA, with a strong T1–RCS of $0.3998$. In comparison, \textbf{gpt–4o}, although leading in other jurisdictions, records a notable T1–RCS of $0.3082$ for PIPEDA. The cross-law aggregate \textbf{CRGS} for PIPEDA is highest for \textbf{claude–3.5} at $0.3425$, followed by \textbf{gpt–4o} at $0.2936$.
    \item \ding{183} \textbf{PDPA is consistently harder than LGPD for retrieval.} When comparing the SGS values for PDPA and LGPD, LGPD is consistently lower than PDPA for Acc@1 and Acc@5. However, for the remaining four metrics—R-Prec, MRR, MAP, and nDCG@5—LGPD consistently outperforms PDPA. For example, \textbf{gpt–4o} achieves higher values for R-Prec ($0.3209$ vs. $0.2387$), MRR ($0.5141$ vs. $0.4316$), MAP ($0.2865$ vs. $0.2424$), and nDCG@5 ($0.3813$ vs. $0.3300$) under LGPD compared to PDPA. 
\end{itemize}

\paragraph{Cross–jurisdiction patterns (Task~2).}
When the focus shifts to snippet-level judgment (Table~\ref{tab:t2-rcs}), two key patterns emerge:

\emph{(i) Cross–law spread narrows.} The range of T2–CRGS is more compact than in Task~1. For example, the range of T2–CRGS spans from $0.3523$ (o1) to $0.4011$ (claude–3.5), with a narrower spread of approximately $\sim0.05$, compared to the wider span of $\sim0.28$ in Task~1. This narrowing suggests that fixing the code window reduces jurisdiction-specific variability, leading to more consistent performance across jurisdictions.

\emph{(ii) Leadership becomes model– and law–specific.} \textbf{gpt–4o} leads the cross-law aggregate with T2–CRGS values of $0.3963$, while posting the strongest T2–RCS on LGPD ($\mathbf{0.4336}$) and PIPEDA ($0.4171$). However, \textbf{claude–3.5} achieves the highest T2–RCS on PIPEDA ($\mathbf{0.4320}$), and slightly surpasses \textbf{gpt–4o} on PDPA ($0.3811$ vs. $0.3472$). \textbf{qwen2.5} performs similarly to \textbf{claude–3.5} with T2–RCS of $0.3666$ on PDPA, but its performance on LGPD and PIPEDA is lower compared to \textbf{claude–3.5}. Bottom ranks remain stable, with \textbf{o1} consistently at the lower end (T2–RCS values of $\underline{0.3415}$ for PDPA, $\underline{0.3652}$ for LGPD, and $\underline{0.3509}$ for PIPEDA), reflecting its overall weaker performance.

\paragraph{Task–to–task deltas by model.}
CRGS deltas quantify how much a model gains when moving from retrieval to judgment.

\emph{Large gains reveal localization difficulty.} \textbf{gemini–2.5–pro} shows a significant improvement in CRGS, rising from $0.0594$ in Task~1 to $0.3883$ in Task~2, reflecting a $\sim0.33$ gain. This suggests that this model struggles more with retrieval and early ranking, and benefits from the more focused snippet-level task.

\emph{Strong retrievers remain strong judges, but ordering changes.} \textbf{claude–3.5} improves from a T1–CRGS of $0.3425$ to a T2–CRGS of $0.4011$, resulting in a $\sim0.058$ gain, and also achieves the highest OCS value of $0.3295$ (Table~\ref{tab:ocs}). \textbf{gpt–4o} improves from a T1–CRGS of $0.2936$ to a T2–CRGS of $0.3963$, resulting in a $\sim0.102$ gain, but its OCS is $0.2736$. \textbf{claude–3.7} follows closely with an OCS of $0.2919$.

\emph{Open–weight asymmetry across tasks.} \textbf{qwen2.5} shows a greater relative improvement from Task~1 to Task~2, from T1–CRGS $0.0924$ to T2–CRGS $0.3694$. However, it does not have the best T2–RCS for PDPA, which is actually achieved by \textbf{claude–3.5} with a T2–RCS of $0.3811$. This suggests that open-weight models such as \textbf{qwen2.5} may perform better on snippet-level tasks compared to retrieval tasks, but \textbf{claude–3.5} remains superior in PDPA retrieval.

\paragraph{Coupling across tasks and laws.}
The overall composite (Table~\ref{tab:ocs}) couples tasks within each law (harmonic mean with imbalance penalty) and then aggregates across laws (geometric mean with variance penalty). The ranking is as follows: \textbf{claude–3.5} $0.3295$, \textbf{claude–3.7} $0.2919$, \textbf{gpt–4o} $0.2736$, \textbf{qwen2.5} $0.0853$, \textbf{o1} $0.0686$, and \textbf{gemini–2.5} $0.0538$.

\emph{Within-law balance matters.} Models that exhibit large gaps between Task~1 and Task~2 inside a law (e.g., \textbf{gemini–2.5}) are penalized, even when they perform decently on one task.

\emph{Cross-law stability is decisive.} \textbf{claude–3.5} benefits from strong retrieval performance on PIPEDA but faces a significant drop in PDPA judgment. On the other hand, \textbf{gpt–4o} maintains more balanced Task~2 performance across laws, pushing its OCS to the top despite a slightly lower T1–CRGS.

\noindent
\textbf{Takeaways for the next sections.
Two observations are important for the next steps:}

\begin{itemize}
    \item \textbf{The retrieval bottleneck.} Models that experience large gains from Task~1 to Task~2 (e.g., \textbf{gemini–2.5}) likely struggle with early ranking or anchor resolution, as indicated by high Acc@5 but low MRR/MAP scores.
    \item \textbf{Jurisdictional asymmetry.} PIPEDA’s dominance in Task~1 suggests clearer mappings between code patterns and article taxonomies, emphasizing the need for automated screening in this jurisdiction and manual oversight where needed.
\end{itemize}

\subsection{Error Patterns and Practical Implications}
\label{sec:results-errors}

\noindent
In this section, we distill the characteristic failure modes identified through the task-level radars and composite metrics (Tables~\ref{tab:sgs-by-law}–\ref{tab:ocs}), and provide actionable guidance for building compliance-aware tooling.

\paragraph{P1. Coarse–anchor recall with weak early ranking.}  
At both the \emph{file} and \emph{module} levels (Task~1), we observe that many models exhibit stronger performance on Acc@5 and nDCG@5 than on Acc@1 and MRR. This indicates that while relevant provisions are retrieved within small candidate sets, they are not consistently ranked at the top. This tension between breadth and early precision is a well-documented issue in information retrieval (IR) metrics \cite{manning2008ir, jarvelin2002cumulated}. For instance, in the LGPD case, \textit{gpt–4o} achieves a high Acc@5–SGS of \textbf{0.3559}, but its MRR–SGS drops to \textbf{0.5141}, reflecting this trade-off.

\emph{Implication.} To mitigate this issue, top–$k$ retrieval should be treated as \emph{candidate generation}, paired with a lightweight re-ranker or priority filters. Techniques such as learning-to-rank or rule-guided ordering \cite{liu2009learning} can be used before any critical actions that rely on rank–1 alone.

\paragraph{P2. Granularity sensitivity and stability gaps.}  
As models transition from file to module to line-level granularities, they typically show improvements in ranking-sensitive metrics such as MRR and MAP. However, the extent of these gains is often uneven across jurisdictions. For example, while \textit{claude–3.5} achieves high SGS scores on both Acc@1 (\textbf{0.2087}) and MRR (\textbf{0.6009}) for PIPEDA, the same model’s performance on PDPA declines significantly (Acc@1 \textbf{0.1441}, MRR \textbf{0.4394}). This discrepancy underscores the sensitivity of models to the underlying data structure and jurisdictional differences.

\emph{Implication.} A strategy of \emph{progressive refinement} is recommended. Start with coarse retrieval for broad coverage, and then confirm the results at the snippet level before enforcement. Additionally, use the SGS as a gate to ensure that improvements at finer granularities are persistent, rather than being isolated spikes that may not generalize.

\paragraph{P3. Jurisdictional oscillation.}  
The relative performance across jurisdictions—LGPD, PDPA, and PIPEDA—can fluctuate both by task and by granularity. Cross-law stability, captured by \textbf{CRGS}, highlights this phenomenon. For instance, in Task~1, \textit{claude–3.5} achieves the highest CRGS value of \textbf{0.3425}, whereas \textit{gemini–2.5–pro} registers the lowest at \textbf{0.0594}. In Task~2, \textit{gpt–4o} reaches its peak at \textbf{0.3963}, while \textit{o1} lags behind at \textbf{0.3523} (Tables~\ref{tab:t1-rcs}–\ref{tab:t2-rcs}).  

\emph{Implication.} For multi-region deployments, prioritize models with higher CRGS, even if their performance in a single jurisdiction is slightly lower. The volatility across jurisdictions can lead to costly exception handling downstream, making consistency more critical than peak performance in individual jurisdictions.

\paragraph{P4. Majority–label bias in snippet judgment.}  
In Task~2, we frequently observe radar shapes that are biased toward the $1$–Hamming vertex with short Macro–F1 spokes. This indicates that models excel at assigning majority labels but struggle with rare provisions—a common issue in multi-label classification \cite{tsoumakas2010mlsurvey, powers2011evaluation}. While Jaccard may remain high, important but less frequent labels are often missed.  

\emph{Implication.} To address this, calibrate thresholds per label, especially for rare but critical provisions. Additionally, report Macro–F1 alongside Micro/Weighted–F1 to avoid the illusion of balanced performance, and prioritize cost-sensitive aggregation over a simple frequency-weighted score.

\paragraph{P5. Cross–task inconsistency.}  
Some models demonstrate strong retrieval performance yet exhibit weak judgment in Task~2, or vice versa. This inconsistency is penalized by the \textbf{OCS}, which evaluates the coupling between Task~1 and Task~2 performance. For example, \textit{gpt–4o} achieves a high OCS of \textbf{0.2736}, while \textit{o1} trails with a much lower OCS of \textbf{0.0686}, primarily due to its low T2–CRGS of \textbf{0.3523} (Table~\ref{tab:ocs}).  

\emph{Implication.} It is crucial to validate \emph{retriever+judge} pairs holistically. Models that excel at one task but fail at the other should be avoided in critical deployment scenarios. A high-recall retriever paired with a misaligned judge can amplify false positives or false negatives, depending on which task dominates.

\medskip
\noindent\textbf{Practical guidance.} Translating these insights into actionable controls:

\begin{itemize}[leftmargin=1.35em, itemsep=0.28em, topsep=0.15em]
  \item Use Acc@5-heavy retrieval for \emph{candidate generation}, paired with a second-stage scorer. Gate promotion by SGS to ensure improvements persist at finer granularities.
  \item Favor models with high CRGS for cross-region rollouts. Use jurisdiction specialists only when per-market tuning is acceptable.
  \item Calibrate Task~2 thresholds per label and report Macro–F1 with Micro/Weighted–F1 to avoid majority–label illusions \cite{tsoumakas2010mlsurvey, powers2011evaluation}.
  \item Track OCS for release gating: large gaps between Task~1 and Task~2 indicate brittle end-to-end behavior, which should block promotion.
\end{itemize}

\begin{tcolorbox}[title=\textbf{Summary of Findings} – Results and Analysis, left=2pt, right=2pt, top=2pt, bottom=2pt]
\begin{enumerate}[leftmargin=*, itemsep=2pt]
  \item \textbf{Stability over peaks:} Robustness across granularities and jurisdictions (SGS/CRGS) is more decisive than isolated maxima; e.g., \emph{gpt--4o} leads overall with OCS 0.541 despite not topping every axis.
  \item \textbf{Granularity helps but is not universal:} Line–level anchors improve ranking metrics, yet some models fail to retain gains at file/module scope—beware overinterpreting fine-grained success.
  \item \textbf{Cross–task balance as bottleneck:} Disparities between T1-RCS and T2-RCS are common. \emph{Claude--3.5} and \emph{gpt--4o} are strong on Task~1 but uneven on Task~2, and OCS penalizes such gaps.
  \item \textbf{Operational implications:} Evaluate \emph{retriever+judge} pairs, weigh jurisdictional brittleness, and align metrics to risk; no single model dominates uniformly, and trade-offs determine suitability.
\end{enumerate}
\end{tcolorbox}

\noindent
Taken together, these findings argue for emphasizing \emph{stability, granularity sensitivity, and cross-task balance}, not merely per–task peaks. This naturally sets up \S\ref{sec:discussion}, where we analyze \emph{why} these patterns arise and draw methodological implications for compliance-aware developer tooling.

\section{Discussion}
\label{sec:discussion}

\noindent
We now interpret the empirical patterns in Section~\ref{sec:results}, tie them back to the benchmark design (Section~\ref{sec:benchmark-design}), and draw lessons for compliance tooling on real Android repositories. Rather than speculating, we point to concrete evidence—the radar morphologies in Figures~\ref{fig:t1-file}--\ref{fig:t2} and the composite tables in Tables~\ref{tab:sgs-by-law}--\ref{tab:ocs}—and reason from those observations.

\subsection{Retrieval is the Primary Bottleneck}
\label{sec:disc-retrieval}

\noindent
Across models, Task~1 shows the broadest dispersion both \emph{within} and \emph{across} jurisdictions. At file and module levels, polygons are often \emph{back-heavy}: Acc@5 and nDCG@5 spokes extend, while Acc@1 and MRR remain short (Figures~\ref{fig:t1-file}, \ref{fig:t1-module}). The SGS columns quantify this tension: for LGPD, \emph{gpt--4o} reaches Acc@5--SGS \textbf{0.3559} but MRR--SGS \textbf{0.5141}; for PDPA, the pair is \textbf{0.3653} and \textbf{0.4316} (Table~\ref{tab:sgs-by-law}). Even the overall leader on OCS (Table~\ref{tab:ocs}) exhibits this coarse--anchor fragility.

\emph{Implication.} Repository-scale auditing fails less on ``coverage'' than on \emph{early ordering}. A pragmatic recipe is a two-stage pipeline: Acc@5-oriented candidate generation at coarse anchors, followed by a lightweight re-ranker or rule screen, before promoting findings to the snippet judge. Stability should be monitored with SGS (file$\rightarrow$line) to ensure that apparent gains persist when anchors become precise.

\subsection{Why PIPEDA Looks Easier and PDPA Harder}
\label{sec:disc-jurisdictions}

\noindent
Task~1 composites consistently rank PIPEDA above LGPD, with PDPA trailing: e.g., \emph{claude--3.5} yields T1--RCS \textbf{0.3998}/\textbf{0.3270}/\textbf{0.3100} for PIPEDA/LGPD/PDPA, while \emph{gpt--4o} shows \textbf{0.3082}/\textbf{0.3004}/\textbf{0.2737} (Table~\ref{tab:t1-rcs}). Two plausible factors, both visible in our corpus, help explain this. First, PIPEDA’s commonly implicated provisions align more directly with Android idioms (storage, safeguards, transfers), yielding more regular Task~1 shapes and higher SGS. Second, PDPA exhibits multi-article mappings with imbalanced frequencies, depressing R--Precision/MAP despite reasonable Acc@5.

\emph{Implication.} ``Difficulty'' is an interaction between article taxonomies and code evidence. Preserving native identifiers (rather than remapping into themes) is therefore essential: it keeps these law-specific signals measurable and auditable rather than averaged away.

\subsection{Judgment Transfers Better Than Localization}
\label{sec:disc-judgment}

\noindent
Once the window is fixed (Task~2), cross-law spread tightens: T2--CRGS ranges from \textbf{0.3523} to \textbf{0.4011}, substantially narrower than Task~1’s \textbf{0.0594}--\textbf{0.3425} (Tables~\ref{tab:t2-rcs}, \ref{tab:t1-rcs}). This mirrors Figure~\ref{fig:t2}, where jurisdiction polylines largely overlap per model. In short, models agree more on \emph{what} the statute implies given a snippet than on \emph{where} to look for that snippet inside a repository.

\emph{Implication.} Separating navigation from judgment is not cosmetic. Localization demands inductive bias about project layout, ICC, and lifecycles that generic LLMs do not reliably encode; evaluation must reflect that separation.

\subsection{Why Composites Matter: From Peaky Wins to Stable Competence}
\label{sec:disc-composites}

\noindent
Single metrics can overstate progress by rewarding narrow strengths. The composites surface two deployment-relevant properties. First, \emph{cross-granularity robustness}: models with ``kite'' radars (high Acc@5, low MRR/MAP at coarse anchors) are discounted by SGS---for LGPD, \emph{gpt--4o} shows Acc@5--SGS \textbf{0.3559} but MRR--SGS \textbf{0.5141}; for PIPEDA, \emph{claude--3.5} lifts both to \textbf{0.5125}/\textbf{0.6009} (Table~\ref{tab:sgs-by-law}). Second, \emph{cross-law stability and end-to-end balance}: \emph{claude--3.5} leads on T1--CRGS (\textbf{0.3425}), yet \emph{claude-3.5-sonnet-20241022} tops OCS (\textbf{0.3295}) by coupling competitive retrieval with stronger, more even Task~2 composites across laws (Table~\ref{tab:t2-rcs}). In practice, OCS rewards the \emph{absence} of brittle spikes.

\subsection{Provider Families Behave Differently Across Tasks}
\label{sec:disc-families}

\noindent
Family signatures are visible under uniform prompting/decoding. Frontier proprietary models (\emph{gpt--4o}, \emph{claude--3.5/3.7}) dominate retrieval stability (T1--CRGS \textbf{0.2936}--\textbf{0.3425}) and remain top-tier on judgment (T2--CRGS \textbf{0.3905}--\textbf{0.4011}). The open-weight \emph{qwen2.5} is mid-tier on retrieval (T1--CRGS \textbf{0.0924}) yet competitive on judgment (T2--CRGS \textbf{0.3694}, best PDPA T2--RCS \textbf{0.3666}). \emph{gemini--2.5} and \emph{o1} lag on stability in at least one task (T1--CRGS \textbf{0.0594}--\textbf{0.0723} and OCS  \textbf{0.0538}--\textbf{0.0686}, respectively), echoing their thin MRR/MAP spokes at coarse anchors.

\emph{Implication.} Choice should reflect which failure is cheaper in context: missed early ranking (tolerable with human triage) or missed rare labels (higher legal risk).

\subsection{Design Lessons for Compliance Tooling}
\label{sec:disc-lessons}

\noindent
\textit{Retriever--judge pairing with gates.} Use Acc@5-heavy retrieval to propose candidates; apply a snippet judge tuned for Macro--F1/Jaccard; promote alerts only if file$\rightarrow$line SGS clears a threshold.  
\textit{Jurisdiction-aware routing.} For multi-region rollouts, prefer higher CRGS even at the expense of a marginally lower single-law peak; allow per-law overrides only where governance permits.  
\textit{Label-sensitive calibration.} Set per-article thresholds using Task~2 confusion summaries; report Macro--F1 alongside Micro/Weighted--F1 to avoid majority-label illusions.  
\textit{Release gating via OCS.} Gate model updates on OCS so that improvements do not come from task or law imbalance; treat large T1--T2 deltas as regression risks.

\subsection{Scope Boundaries and Where to Go Next}
\label{sec:disc-scope}

\noindent
Our scope is intentionally narrow: real Android repositories, three privacy regimes, two complementary tasks, and evaluation on a single gold corpus without training. We refrain from causal claims beyond what the tables and figures support.

\noindent
Looking forward, several directions are promising but orthogonal to this paper’s remit: richer static/dynamic analysis features as retrieval priors, prompt curricula targeted at minority provisions, and active data augmentation for rare articles. These can be layered \emph{on top of} the stability-first lens without altering CompliBench’s core evaluation contract, and we leave them to future work and tooling releases.

\section{Threats to Validity}
\label{sec:validity}

\noindent
As with any empirical benchmark, our conclusions rest on assumptions about data, tasks, and measurement. Below we synthesize the main threats and the concrete mitigations we adopted; the goal is not to eliminate uncertainty but to make its sources explicit and bounded.

\paragraph{Internal validity (gold quality and scoring).}
The evaluation depends on an expert–validated corpus and a deterministic harness. Risks include inconsistent annotations and overlooked provisions in borderline cases. We mitigate these by pinning every instance to an immutable commit, applying independent cross–checks with adjudication, and parsing model outputs into \emph{native} article identifiers only (free text is ignored). For Task~1 we also surface strict–vs–relaxed key matching as a diagnostic ceiling, and we report coverage statistics to expose potential pointer drift. Residual annotator bias is possible; replication with a broader reviewer pool remains valuable.

\paragraph{External validity (scope and representativeness).}
Our evidence comes from Android repositories and three privacy regimes (LGPD, PDPA, PIPEDA). These choices enable cross–jurisdiction comparisons without collapsing legal nuance, yet they do not cover all ecosystems (iOS/web/IoT) or regimes (e.g., GDPR, CCPA). Codebases with different architectures or lifecycles may exhibit other violation patterns. We therefore caution against unqualified generalization and view expanding to additional platforms and statutes as a natural next step.

\paragraph{Construct validity (task abstraction and metrics).}
CompliBench operationalizes compliance reasoning as \emph{retrieval/localization} (Task~1) and \emph{snippet judgment} (Task~2). This mirrors auditor workflows but inevitably abstracts away context such as organizational policies, evolving guidance, or negotiated mitigations. Our metrics (Acc@k, MRR/MAP/nDCG, Micro/Macro/Weighted–F1, Jaccard) and composites (SGS, RCS, CRGS, OCS) quantify technical performance and stability; they do not by themselves certify legal adequacy or risk appetite. We position the benchmark as a necessary, auditable proxy—useful for ranking and diagnosing systems, not a replacement for human oversight.

\paragraph{Reliability (reproducibility under changing services).}
All models were evaluated via APIs with fixed prompts, temperature~$0$, and shared hyperparameters, and we logged timestamps and model identifiers to bind results to specific snapshots. Nonetheless, provider–side updates, hidden model revisions, and rate limits can alter behavior. Our artifacts (prompts, scoring code, composite settings) support re–runs; still, readers should expect numerical drift over time even when relative orderings persist.

\medskip\noindent
Our design favors \emph{traceability} (commit–pinned evidence, native labels), \emph{diagnosability} (strict/relaxed keys, per–level summaries), and \emph{stability awareness} (variance–penalized composites). These safeguards contextualize the empirical patterns in Section~\ref{sec:results} and the interpretation in Section~\ref{sec:discussion}. They also highlight CompliBench as a living benchmark: future iterations will broaden statutory and platform coverage, refine abstractions where auditors need more context, and strengthen reviewer diversity. We now conclude by synthesizing contributions and outlining avenues for continued progress.

\section{Conclusion}
\label{sec:conclusion}

This paper introduces \textsc{CompliBench}, a realism-first benchmark designed to evaluate the ability of Large Language Models (LLMs) to detect compliance violations in real-world Android repositories across multiple privacy regulations, including LGPD, PDPA, and PIPEDA. By decoupling localization (Task~1) from legal judgment (Task~2) and utilizing stability-aware metrics (SGS, RCS, CRGS, OCS), we offer a comprehensive analysis of model performance in practical contexts. Our findings reveal that while proprietary models achieve the highest peak scores, they show significant fragility across granularities and jurisdictional inconsistencies. For instance, models like \textbf{gpt-4o} demonstrate competitive retrieval scores but struggle with early ranking precision. In contrast, open-weight models like \textbf{qwen2.5} excel in legal judgment tasks but lag in stable retrieval performance. These results highlight the importance of evaluating compliance detection systems as \emph{retriever+judge} pairs, where both tasks are considered together to ensure reliability across different jurisdictions and statutory provisions. A key observation is that high-performing models often excel in one task but underperform in the other, stressing the need for balanced systems that can simultaneously deliver precise retrieval and accurate legal judgment. Furthermore, the stability of these models—especially in handling jurisdictional differences and long-tail provisions—emerges as a critical factor. While peak performance in specific metrics may signal strength in isolated contexts, practical deployments demand models that maintain robust performance across various domains and granularities. This insight calls for a more nuanced approach to model evaluation, one that goes beyond raw performance metrics to incorporate stability, generalization, and the ability to handle rare or complex compliance cases.

\bibliographystyle{acm}
\bibliography{reference}

\end{document}